\documentclass[11pt]{article}

\usepackage[a4paper, margin=0.8in]{geometry}

\usepackage{amsmath, amsthm, amsfonts, rotating}

\usepackage{bm}
\usepackage{scalerel}

\usepackage[T1]{fontenc}              
\usepackage[utf8]{inputenc}         
\usepackage{natbib}   
\usepackage[english]{babel}

\usepackage{amsthm,amsmath,amsfonts,amssymb} 
\usepackage{etoolbox}

\makeatletter
\patchcmd{\@makecaption}
  {\parbox}
  {\advance\@tempdima-\fontdimen2} 
  {}{}
\makeatother 

\usepackage{graphicx}                 

\usepackage{listings}                 
\usepackage{tikz}
\usepackage{tikz-3dplot}
\usetikzlibrary{math} 
\usetikzlibrary{positioning}

\usepackage{mathtools}
\usepackage{enumerate}

\usepackage{epstopdf}
\usepackage{bbm}
\usepackage{placeins}
          
\usepackage{standalone}
\usepackage{url}
	
\usepackage{soul}

\theoremstyle{plain}

\theoremstyle{definition}

\theoremstyle{remark}

\usepackage{booktabs}
\usepackage{array}
\usepackage{multirow}
\usepackage{wrapfig}
\usepackage{float}
\usepackage{colortbl}
\usepackage{pdflscape}
\usepackage{tabu}
\usepackage{threeparttable}
\usepackage{threeparttablex}
\usepackage[normalem]{ulem}
\usepackage[utf8]{inputenc}
\usepackage{makecell}
\usepackage{verbatim}
\usepackage[symbol]{footmisc}

\newcommand{\bV}{\mbox{\boldmath $V$}}

\newcommand{\bh}{\mbox{\boldmath $h$}}

\newcommand{\bms}{\mbox{\boldmath $s$}}

\newcommand{\bbeta}{\mbox{\boldmath $\beta$}}

\title{Small Area Estimation of Education Levels in Low- and Middle-Income Countries}
\date{\vspace{-5ex}}

\author{Yunhan Wu$^1$, Ameer Dharamshi$^1$, Jon Wakefield$^{1,2}$\\
\vspace{0.2em}\\
$^1$Department of Biostatistics, University of Washington, Seattle, USA\\ 
$^2$Department of Statistics, University of Washington, Seattle, USA \phantom{te}}

\begin{document}

        \maketitle

\begin{abstract}
Education is a key driver of social and economic mobility, yet disparities in attainment persist, particularly in low- and middle-income countries (LMICs). Existing indicators, such as mean years of schooling for adults aged 25 and older (MYS25) and expected years of schooling (EYS), offer a snapshot of an educational system, but lack either cohort-specific or temporal granularity. To address these limitations, we introduce the ultimate years of schooling (UYS)—a birth cohort-based metric targeting the final educational attainment of any individual cohort, including those with ongoing schooling trajectories. As with many attainment indicators, we propose to estimate UYS with cross-sectional household surveys. However, for younger cohorts, estimation fails, because these individuals are right-censored leading to severe downwards bias. To correct for this, we propose to re-frame educational attainment as a time-to-event process and deploy discrete-time survival models that explicitly account for censoring in the observations. At the national level, we estimate the parameters of the model using survey-weighted logistic regression, while for finer spatial resolutions, where sample sizes are smaller, we embed the discrete-time survival model within a Bayesian spatiotemporal framework to improve stability and precision. Applying our proposed methods to data from the 2022 Tanzania Demographic and Health Surveys, we estimate female educational trajectories corrected for censoring biases, and reveal substantial subnational disparities. By providing a dynamic, bias-corrected, and spatially disaggregated measure, our approach enhances education monitoring; it equips policymakers and researchers with a more precise tool for monitoring current progress towards education goals, and for designing future targeted policy interventions in LMICs.

\bigskip
\noindent
\textbf{Keywords}: Educational attainment; Bayesian smoothing; Spatial models; Time series models; Censoring; Survival modeling; Small area estimation
\end{abstract}

\section{Introduction}

Universal access to high-quality education is a central pillar of the global development agenda. Indeed, it is well understood that the level of one's education is a catalyst for social and economic mobility; greater educational attainment, defined as the number of years of schooling completed, is a direct determinant of a wide swath of health outcomes, longevity, fertility, economic growth, and overall societal progress and well-being \citep{barro1993international, manda2005age, meara2008gap, lutz2008demography, fuchs2010education}. In recognition of the important role of education, the global community has committed to {``}ensure inclusive and equitable quality education and promote lifelong learning opportunities for all{''}, and has enshrined this objective in the fourth Sustainable Development Goal (SDG 4) \citep{SDG4}.

In recent decades, there has been substantial progress in expanding access to education and increasing attainment. Globally, the percetage of children and youth completing primary education (with at most 5 years delay) increased from 82.8\% in 2010 to 87.9\% in 2023 \citep{UIS}. Yet major disparities persist, particularly in low- and middle-income countries (LMICs). Subsetting to Sub-Saharan Africa, over the same time period the primary completion rate grew from 55.9\% to 67.3\% \citep{UIS}.

As we look forward, there is a pressing need for high-quality educational attainment data extending over long periods of time and disaggregated at fine spatial resolutions. Such data is essential for monitoring progress towards universal education, informing data-adaptive education policies at local, national, and international levels, and to study the downstream impacts of changing education dynamics on other societal outcomes. Unfortunately, this data is not readily available. Full population enumerations are scarce, especially in LMICs, and even in cases where administrative data systems do exist, they often lack data on attainment or suffer from incomplete coverage \citep{dharamshi2022bayesian}. 

In modern practice, monitoring of educational attainment instead relies heavily on censuses and on high-quality household survey data such as those produced by the Demographic and Health Surveys (DHS), Multiple Indicator Cluster Survey (MICS), and many other local survey programs. In this paper, we focus on DHS data due to its extensive coverage, with over 300 surveys conducted across 90+ countries \citep{croft2018guide}, making it one of the most comprehensive sources of health and demographic information in LMICs, though the discussions in the remainder of this paper apply broadly. Briefly, the DHS program is a series of household surveys that collect extensive data on health, demographic, and social indicators. These data are used by national and international organizations to monitor progress toward development targets within each country. The DHS asks respondents to provide detailed information on their education histories; most notably in our context, they ask about the highest level of education completed and the number of years completed in the following level. Together, these questions lead to a measure of the total number of years of schooling completed; that is, they offer reliable information which can be used to fit statistical models of educational attainment.

While the potential utility of household survey data to monitor educational attainment is clear, several important open questions remain. At present, there is a lack of consensus on the optimal way to define and report attainment. The Human Development Index (HDI) uses two metrics: the expected years of schooling (EYS) for children of school-entry age computed with respect to prevailing age-specific enrollment rates, and the mean years of schooling for adults aged 25 years and older (MYS25). While the former provides an outlook on the future educational levels of the population (assuming no progress or backsliding), the latter reflects current educational attainment \citep{smits2019subnational}. These indicators offer a snapshot of an educational system and are therefore useful as components of a composite index like the HDI, but are limited for other purposes. For example, if our interest is in examining the association between maternal educational attainment and fertility outcomes, MYS25 lacks the temporal and cohort-specific granularity necessary to adequately answer the question of interest.

To better capture changes in educational attainment across generations, we propose a new metric: the birth cohort-specific years of schooling. The birth cohort-specific years of schooling is defined as the average years of schooling that individuals in a specific birth cohort are expected to complete, accounting for ongoing educational trajectories. As it represents the ultimate attainment level of a birth-cohort, we will henceforth refer to this metric as the ultimate years of schooling (UYS). 

UYS is quite different from the two traditional metrics. It is defined with respect to individual cohorts, which are the natural structure of educational trajectories. This is in contrast to EYS, which estimates the number of years a child of school-entry age is expected to spend in school, assuming that current (calendar year) age-specific enrollment rates remain constant (an assumption that may not be realistic in settings with evolving education contexts), and with MYS25 which is an aggregate measure. Consider again the example of maternal educational attainment and fertility outcomes. UYS is dynamic, it provides a natural means of exploring the covariation between cohort-specific attainment and cohort-specific fertility outcomes, a feat that neither of the other indicators can accomplish.

In this paper, we propose a framework for estimating time series of UYS in LMIC contexts in small subnational areas. There are two primary methodological challenges that must be addressed: the incomplete nature of data on recent cohorts that have yet to fully complete their schooling, and the challenges associated with small area estimation (SAE) using household survey data.

The first challenge arises because DHS surveys capture educational histories for both individuals still in school and those who have completed their education. If we simply compute the weighted average of years of schooling for younger cohorts, the estimates will suffer from potentially severe downwards bias as those still in school will contribute a truncated number of years. While survey weights account for the complex sampling design, they do not address this truncation issue. This risk is particularly acute in LMICs as delayed schooling, often by several years, is common \citep{GEM2024}. MYS25 sidesteps this issue with an age threshold of 25 years. Indeed, the justification for the age threshold is that 25 years marks the typical completion of formal education, therefore eliminating the issue of incomplete cohorts. The cost of the thresholding approach is timeliness as it renders MYS25 a lagging indicator of education policies in the sense that we must wait until a cohort has fully aged through the educational system before they contribute any information to MYS25 monitoring. \cite{dharamshi2022bayesian} faced a similar challenge in their estimation of school completion rates, a measure of timely attainment. They proposed to correct for the downwards bias by parameterizing and estimating it with a piecewise linear function. 

\begin{figure} [!ht]
    \centering
    \includegraphics[clip, trim=0cm 0cm 0cm 0cm, width=0.98\linewidth]{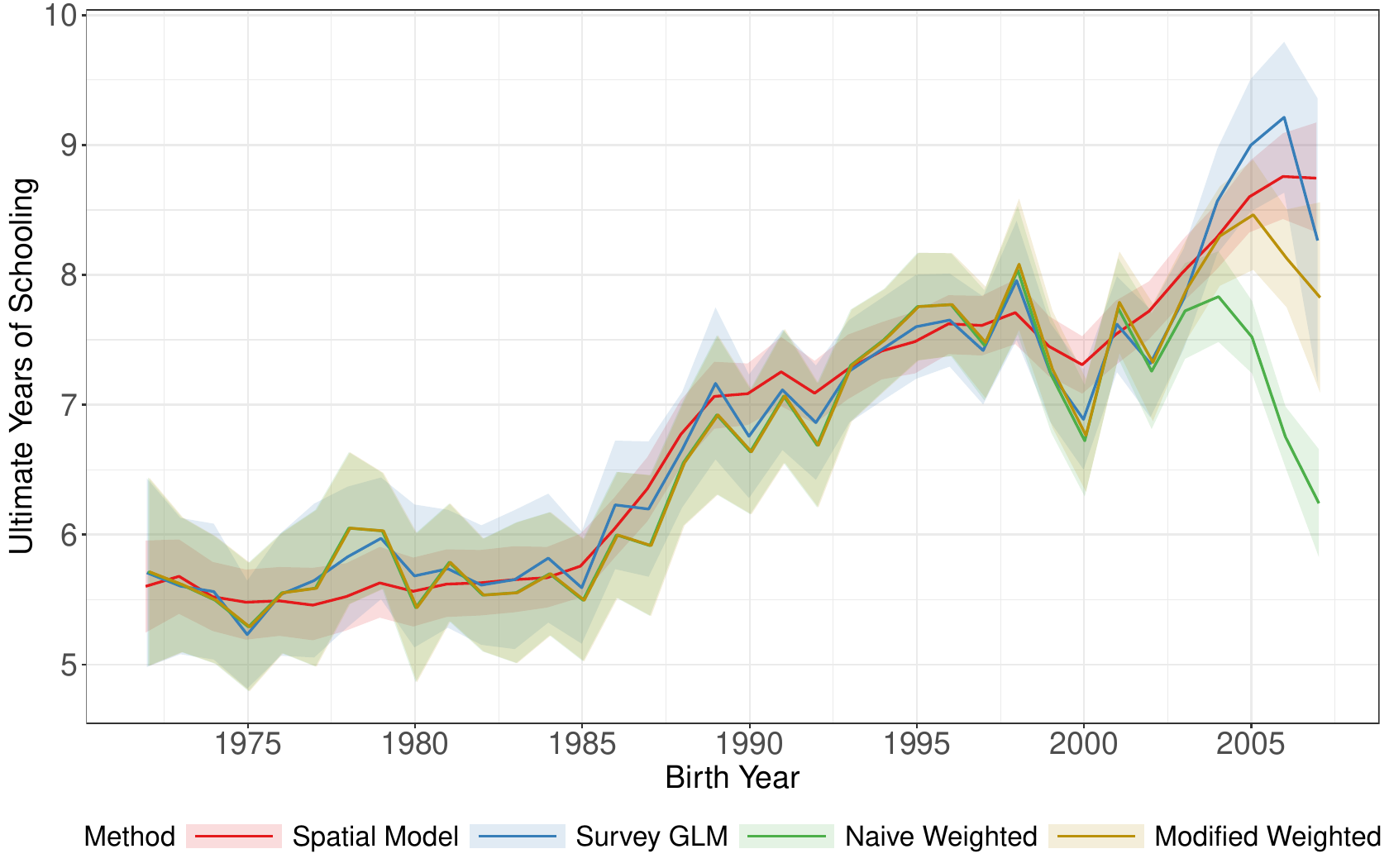}
    \caption{Estimated UYS by birth year for females in the 2022 Tanzania DHS, across methods to be introduced later.
}
    \label{fig:natl-overall-educ-yrs-TZA-female}
\end{figure}

We take a different approach that neither needlessly discards data, nor relies on strong parametric assumptions on the form of the bias. We propose to re-frame the concept of educational attainment as a time-to-event outcome where the event of interest is exit from the educational system. From this viewpoint, young cohorts are not incomplete; rather, they are \emph{right-censored}. Given that years of schooling are recorded on a discrete scale, we estimate UYS using an individual-level discrete-time survival model \citep{brown1975on, allison1982discrete, singer1993it}. This approach is both intuitively and technically appealing: on the former, the cohort-specific hazard rates are understood as the cohort-specific risk of dropping out of school given currently in school, a quantity of considerable substantive interest; and on the latter, a proportional odds assumption enables estimation of the complete trajectories of youth cohorts.

The impact of right-censoring and its potential to bias UYS estimates downward is illustrated in Figure \ref{fig:natl-overall-educ-yrs-TZA-female}. In the 2022 Tanzania DHS, over 60\% of 15-year-old females were still in school, meaning their ultimate educational attainment remains unobserved and thus right-censored. In contrast, among 20-year-olds, only 6\% remained enrolled, indicating minimal censoring for this cohort. Figure \ref{fig:natl-overall-educ-yrs-TZA-female} presents results from methods to be introduced later, highlighting differences in how they handle right-censored education attainment data. The naive weighted estimate (green line) declines sharply for recent cohorts as it ignores right-censoring entirely. The modified weighted estimate (yellow line) partially adjusts for this but remains limited.  In contrast, the model-based estimates (red and blue lines) account for ongoing education, offering a complete picture of educational trajectories. For older cohorts (25+ years), all estimates converge as their full schooling histories are observed. The methods behind these estimates will be detailed in subsequent sections.

To estimate the parameters of the discrete-time survival models, we propose two strategies. When the targets of inference are large areas for which representative samples are available (i.e., national), we first exploit a connection between discrete-time survival models and logistic regression \citep{singer1993it}. This reparameterization allows standard methods implemented in the \texttt{survey} R package \citep{survey} to be used. When the targets are smaller areas (such as the first and second principal administrative divisions, henceforth referred to as Admin-1 and Admin-2, respectively) which suffer from low sample size problems, we embed the discrete-time survival construction within a Bayesian spatiotemporal model that allows information to pool across time and space, thereby stabilizing estimation \citep{wakefield2019estimating}. 

We view the extension to small subnational areas as a meaningful contribution to the education modelling literature. The education monitoring framework has thus far largely focused on national level estimation and cross-country comparisons. \cite{delprato2024spatial} advocate for a shift towards subnational monitoring, providing qualitative and quantitative evidence of substantial subnational variation in attainment related indicators across sub-Saharan Africa. These local disparities largely stem from the fact that operationally, the delivery and context of educational programs is a highly local affair. Educational inhibitors such as distance to the nearest school \citep{macharia2023modelling}, child labour practices \citep{lewin2009access}, and early marriage \citep{delprato2022zones} can be highly heterogeneous within countries, and therefore subnational data are needed to identify disparities within countries, and guide polices to counteract these gaps. A notable exception to the national focus are the estimates of subnational attainment for 15--19 year-olds, 20--24 year-olds, and 15--49 year-olds across Africa produced by \cite{graetz2018mapping}. They, however, focus on aggregate indicators rather than individual birth cohorts, and target attainment at the time of survey rather than ultimate attainment as we do here i.e.,~they do not correct the bias due to incomplete educational trajectories for youth populations. 

By employing statistical models that account for the complexities of incomplete education at the time of data collection, we are able to reliably estimate the full educational trajectories of individuals across birth cohorts and subnational geographies. Our approach is accurate and intuitive; it is fundamentally rooted in the natural pathway of student cohorts progressing through an educational system. This structure offers researchers, policymakes, and practitioners a clearer understanding of educational progress in LMICs both through the final UYS estimates, as well as through useful auxiliary parameters that map each cohort's education journey.

The remainder of this paper is organized as follows. In Section \ref{sec:method}, we describe in detail the issues with weighted estimation in this context, and propose a set of increasingly complex discrete-time survival methods to estimate UYS. We then apply our proposed models in simulation in Section \ref{sec:sims}, and to the Tanzania DHS data for females in Section \ref{sec:results}. We conclude with a discussion of our method and findings in Section \ref{sec:discussion}. Code to reproduce all results in this paper is available at {\tt https://github.com/wu-thomas/UYS-SAE}.

\section{Method}
\label{sec:method}

We first describe a national model, focusing on the birth cohort-specific censoring feature of the method, before turning to the spatial domain and temporal aspects. 

\subsection{Naive Weighted Estimation}
\label{sec:direct-unadj}

DHS surveys record the total years of schooling completed by each respondent up to the time of the survey. For the \( j \)-th individual, let \( t_j \) denote the years completed.
We aim to estimate the average years of education for individuals within birth cohorts indexed by \( b \).  To simply notation, in the following we begin by omitting additional indices for space and urban/rural status. Let \( S_b \) denote the set of individuals sampled from birth cohort \( b \), with \( w_j \) being the associated design weight and \( t_j \) the response for individual \( j \).

We begin with a straightforward approach called \textit{weighted estimation}, which produces an estimate for a specific cohort using only the response data from that cohort. A common example is the design-based weighted estimator \citep{hajek1971discussion}. The weighted estimate of the average years of education for domain \( b \) is calculated as:
\begin{equation}
\label{eq:svy-weighted-educ-year-unadj}
    \hat{\mu}_b = \frac{\sum_{j \in S_b} w_{j} t_{j}}{\sum_{j \in S_b} w_{j}}.
\end{equation}
We emphasize that this estimator relies exclusively on data from the sampled individuals within the cohort without considering the right-censoring, and thus we term it \textbf{naive weighted estimation}. The variance of $\hat{\mu}_b$ can be easily estimated with standard software implementations for design-based estimators.

A disqualifying limitation of the naive weighted estimation approach in our context is that it fails to account for right-censoring in educational trajectories. For recent birth cohorts, notably those aged 15--19 at the time of the survey, observed years of education often do not reflect final educational attainment (i.e., UYS), as many individuals are still enrolled in school. Treating these right-censored observations as complete effectively assumes that all individuals have finished their education by the survey date, resulting in systematic underestimation for younger cohorts.

Furthermore, weighted estimation relies solely on data from the specific subpopulation under analysis, resulting in high uncertainty and instability when sample sizes are small. This issue will become particularly pronounced when we consider subnational regions stratified by birth cohorts.

These limitations necessitate more sophisticated methods. Here we propose a family of discrete-time survival models with smoothing over time and space. These models not only address right-censoring appropriately, but the introduction of random effects produce substantially more stable estimates in data-sparse settings. The details of these methods are discussed next.

\subsection{Discrete-time Survival Models}

To accurately estimate UYS, we require models capable of addressing two critical challenges: 1.~accounting for right-censoring in educational trajectories, and 2.~providing precise estimates for specific small domains, including those defined by birth cohort and fine-scale spatial regions.

An intuitive and flexible approach is to conceptualize years of education as the duration from school enrollment until completion or discontinuation. This perspective allows us to frame educational attainment as a time-to-event outcome, drawing on methods from survival analysis \citep{willett1991whether}. Here, the event of interest is cessation of schooling, with right-censoring naturally occurring at the survey date for individuals still enrolled. This analytical framework allows us to directly link the probability of continuing education with factors such as grade level, birth cohort, and geographic region. Furthermore, embedding the model within a Bayesian hierarchical structure facilitates model-based inference, accommodating complex dependencies and more flexible modeling strategies.

In educational settings, time is typically measured discretely, and in DHS surveys, the years of schooling is measured in discrete units of years. Thus, we adopt a discrete-time survival model, specifically, a continuation ratio model \citep{singer1993it} to enable stepwise estimation of the probability that an individual advances to the next grade level within the educational system. Adopting this framework requires two simplifying assumptions on the structure of educational trajectories: at each time step there are only two options, progression or discontinuation (i.e., students do not repeat grades), and after discontinuing education, students do not re-enter the educational system. While repetition and re-entry are important phenomena in their own right, they are difficult to identity from DHS data alone, and so we defer their study to future work. 

Formally, we partition the education history of each respondent into discrete grade intervals, \( x_0 = [0, 1), x_1 = [1, 2), \dots, x_{K-1} = [K-1, K) \), where \( K \) represents the highest recorded years of education in the sample and is assumed to be the maximum possible duration of schooling. Grade interval $x_0$ is less than 1 year of education and accounts for respondents with no formal education. Grade interval $x_1$ is one year completed but less than 2 years, and so on for $x_k$, $k=2,\dots,K$. Define the random variable $T$ to be the number of years \textit{ultimately} completed, with $T=k$ corresponding to the respondent stopping during grade interval $x_k$, i.e., completing $k-1$ years of education. The discrete-time hazard \( h_{k} \) is defined as the conditional probability that a respondent discontinues their education during grade interval \( x_k \), given that they have completed up to \( k \) years of schooling: 

\[
h_{k} = P(T = k ~|~ T \ge k).
\]
In this section we do not consider regions or multiple time periods; consequently, the hazards are not indexed by space or calendar time. The survival function $ S(t)=P(T \geq t) $  represents the probability that an individual completes at least \( t \) years of education for $t \in [0,K]$. It is expressed as:
\begin{equation} \label{eq:educ-year-surv}
S(t) = \prod_{k=0}^{t-1} (1 - h_k), \quad \quad  1 \le t \le K
\end{equation}
where $S(0)=1 $ by definition.

The target of estimation, the ultimate years of schooling (UYS) \( \mu \), can be derived from the survival function as,
\begin{equation}\label{eq:educ-year-from-surv}
\mu = E(T) = \sum_{t=1}^{K} S(t)= f(\mbox{\boldmath $h$}),
\end{equation}
where \( \bh = \{h_k,k=1,\dots,K\} \) are the hazards.

To estimate \(\bh\), we derive the log-likelihood of the observed data \citep{suresh2022survival}. For the \( j \)-th individual with observed years of education \( t_j \), their data contribute to the grade intervals \( x_0, x_1, \ldots, x_{t_j-1} \). Let $\delta_j=0/1$ be the indicator for censored/not censored. If the individual is not censored (\( \delta_j = 1 \)), they also contribute to the interval \( x_{t_j} = [t_j, t_j+1) \) where they discontinued (or completed) their education. Subject $j$ does not contribute to the likelihood of intervals beyond \( x_{t_j} \).

The log-likelihood for the \( j \)-th individual can be expressed as: 
\begin{align*}
\log L_j &= 
\delta_j \log P(T_j = t_j) + (1 - \delta_j) \log P(T_j \ge t_j) \\
&= \delta_j \left( \log h_{jt_j} + \sum_{k=0}^{t_j-1} \log (1 - h_{jk}) \right)
+ (1 - \delta_j) \sum_{k=0}^{t_j-1} \log (1 - h_{jk}) \\
&= \sum_{k=0}^{t_j-1} \log (1 - h_{jk}) + \delta_j \log h_{jt_j}.
\end{align*}
To simplify notation, we introduce an event history indicator:
\begin{equation}
\label{eq:event-history-indicator}
Z_{jk} = \mathbb{I}(T_j \in x_k),
\end{equation}
so that \( z_{jk} = 1 \) if the individual discontinues education in interval \( x_k \), and \( z_{jk} = 0 \) otherwise. For censored individuals, \( z_{jk} = 0 \) for all \( k \leq t_j \). Using this indicator, education history can be formulated into a series of Bernoulli outcomes based on grade intervals:
\begin{equation}
\label{eq:bernoulli-parameterization}
\log L_j =\sum_{k=0}^{t_j-(1-\delta_j)} \left[
z_{jk} \log(h_{jk}) + (1 - z_{jk}) \log(1 - h_{jk})
\right],
\end{equation}

which is the log-likelihood of the product of all the $Z_{jk}\sim\text{Bern}(h_{jk})$ random variables. This reformulation explicitly incorporates right-censoring and models the discontinuation probabilities for each interval using the hazard rates \( h_{jk} \). Such a Bernoulli parametrization facilitates linking hazard rates to individual or domain specific features. 

The following sections explore methods for estimating the hazards \( \bh \), constructing the survival function \( S(t) \), and deriving years of education, \( \mu \). These methods impose structures on underlying patterns in the data, such as the effects of birth cohort and geographic region, to estimate full educational trajectories. By embedding these assumptions within a flexible framework, such as generalized linear models or Bayesian hierarchical models, we can obtain more precise estimates, particularly in data-sparse settings.

\subsubsection{Modified Weighted Estimation}
\label{sec:direct-modified}

The naive weighted estimation method defined in Equation (\ref{eq:svy-weighted-educ-year-unadj}) incorrectly treats individuals experiencing right-censoring as having dropped out of school. To 
mitigate this issue, 
we suggest a \textbf{modified weighted estimation} approach that estimates the hazards $\bh$ and transforms them into years of education using Equation (\ref{eq:educ-year-from-surv}). This approach provides  design-unbiased hazard estimates.

Specifically, under the parameterization specified in Equation (\ref{eq:bernoulli-parameterization}), the survey weighted hazard estimate for birth cohort \( b \) at grade \( k \) is:
\begin{equation*}
\label{eq:svy-weighted-hazard}
    \hat{h}_{b,k} = \frac{\sum_{j \in S_{b,k}} w_{j} Z_{jk}}{\sum_{j \in S_{b,k}} w_{j}},
\end{equation*}
where \( S_{b,k} \) is the risk set consisting of individuals in cohort $b$ who have completed at least \( k \) years of education and are \textbf{not} censored at grade \( k \), and \( Z_{jk} \) indicates whether individual \( j \) discontinued education during grade interval \( [k, k+1) \). The associated design-based variance of the hazards, \( \hat{\bV}(\hat{\bh}) \), can be directly obtained from standard software implementation such as the \textit{survey} package in {\tt R} \citep{survey}. By applying the delta method to the transformation in Equation (\ref{eq:educ-year-from-surv}), we can derive estimates for years of education \( \hat{\mu}_b \) and its asymptotic variance \( \hat{\bV}(\hat{\mu}_b) \). Details of this derivation are in Section \ref{sec:supp-derive-delta-method} of the supplemental materials.

While this modification improves upon the naive weighted estimation method, we still cannot obtain an unbiased estimate of the mean in Equation (\ref{eq:educ-year-from-surv}) for the most recent birth cohorts due to missing observations at higher grades. Specifically, the survivor function cannot be estimated beyond the highest grade observed in the data. For instance, a 15-year-old respondent may have completed at most nine years of education, leaving hazards \( \{h_k\} \) undefined for \( k > 9 \). One approach is to assume that the survivor function \( S(t) \) drops to zero beyond the maximum observed grade, which we adopt here. Consequently, the survival curve is artificially truncated at this limit, causing the estimated years of education to resemble a restricted mean survival time (RMST) \citep{kalbfleisch2002statistical} with an  incorrectly bounded upper limit, leading to a downward-biased estimate for UYS.

The following sections resolve this limitation by modeling hazards as smooth functions of calendar time (across birth cohorts) and years of education (within cohorts). This approach allows us to infer missing education times by assuming hazard rates vary gradually both within and across cohorts.

\subsubsection{Survey-Weighted Generalized Linear Model}
\label{sec:survey-GLM}

To address the limitations of the weighted estimation approaches, we propose an alternative method that incorporates assumptions and borrows information across birth cohorts, enabling the estimation of complete educational trajectories, \emph{including recent birth cohorts}.

We adopt a continuation-odds model parameterization based on event history indicators defined in Equation (\ref{eq:event-history-indicator}) and the likelihood in Equation (\ref{eq:bernoulli-parameterization}). For a given grade \( k \) and cohort \( b \), the logit of the hazard, representing the probability of discontinuing, is modeled as:
\begin{equation}
\label{eq:svy-weighted-GLM-link}
\log \left(\frac{h_{b,k}}{1-h_{b,k}}\right) = \beta_k + \gamma_b,
\end{equation}
where \( \beta_k \) is a grade-specific intercept, capturing the baseline discontinuation odds for each grade, and \( \gamma_b \) represents the cohort effect relative to a designated reference cohort $b_0$, for which we set $\gamma_{b_0}=0$. The effects $\gamma_b$ account for cohort-specific variations in educational trajectories and we include them as fixed effects in this model; in the model introduced in the next section, we will apply a smoothing prior to these effects.

The model is fitted using a survey-weighted generalized linear model, which we term the \textbf{survey GLM}, via {\tt svyglm} in the {\tt survey} package \citep{survey} to accommodate the sampling design of the survey data (e.g., stratified two-stage unequal probability cluster sampling in the case of DHS data). The weights ensure design-consistent parameter estimates and enable the computation of design-based variances \citep{binder1983on}.

By formulating the model with different reference domains,  we simplify the derivation of the asymptotic distribution of \( \hat{\mu}_b \), the estimated mean years of schooling for cohort \( b \). Using the delta method, this distribution can be obtained solely from the design-consistent estimates of the baseline hazards \( \hat{\bbeta} \) and their covariance matrix \( \hat{\Sigma} \), without requiring the inclusion of the cohort-specific effects \( \gamma_b \). Further details on this formulation and the delta method derivation are provided in Section \ref{sec:supp-derive-delta-method} of the supplemental materials.

As we are also interested in within-country variation in UYS, we extend (\ref{eq:svy-weighted-GLM-link}) by introducing a fixed effect \(\alpha_i\) for spatial heterogeneity, where \( i = 1, \dots, n \) indexes the \( n \) subnational areas. The extended model is:  

\begin{equation}
\label{eq:svy-weighted-GLM-link2}
\log \left(\frac{h_{b,i,k}}{1-h_{b,i,k}}\right) = \beta_k + \gamma_b + \alpha_i,
\end{equation}

where \( h_{b,i,k} \) is the risk of leaving school for an individual in cohort \( b \), area \( i \), and school year \( k \), with $\alpha_{i_0}=0$ for the reference area ${i_0}$.

The key assumption underlying the survey GLM is the proportional odds assumption, which implies that the logit hazards of dropping out are parallel across birth cohorts and areas. We consider the latter assumption appropriate for modeling grade-to-grade dropout rates across areas because educational systems within a country often share structural similarities, such as national curricula, enrollment practices, and grade progression policies \citep{unicef2019education}. Exploratory plots using data from the 2022 Tanzania DHS in Section \ref{sec:supp-prop-odds-assumption} of the supplemental materials also supports this assumption. 

\subsubsection{Model-based Unit-level Inference}
\label{sec:unit-level-model}
When moving to finer spatial and temporal resolutions, such as yearly birth cohorts and Admin-2 regions, data sparsity becomes a critical issue; in the extreme case, certain domains may lack observations entirely, rendering the corresponding $\alpha_i$ in Equation \eqref{eq:svy-weighted-GLM-link2} unidentifiable. To address this challenge, we model spatial and temporal dependencies, borrowing information across both regions and birth cohorts to enhance estimation precision.

Small area estimation (SAE) formulations that directly model individual outcomes are known as \textit{unit-level models}. These models provide greater flexibility in capturing dependencies and interactions, enabling for a more nuanced representation of spatiotemporal variation ---a capability that the GLM framework lacks. Given their focus on subnational estimates, we refer to these models as \textbf{spatial models}.

Let \( c \) denote the cluster index and \( i[\mbox{\boldmath $s_c$}] \) the administrative area in which cluster \( c \) is located. We decompose each individual's education history into Bernoulli outcomes, 
which are then grouped by birth cohort and grade. For cluster \( c \) and birth cohort \( b \), let \( n_{b,c,k} \) represent the number of individuals attaining at least \( k \) years of education, and \( Y_{b,c,k} \) denote the number who did not advance to the next grade. Following Equation (\ref{eq:event-history-indicator}) and (\ref{eq:bernoulli-parameterization}):
\[
Y_{b,c,k} = \sum_{j \in c} Z_{b,k,c,j},
\]
where \( Z_{b,k,c,j} \) is the indicator for whether individual \( j \) discontinued education within grade interval \( [k,k+1) \). 

To account for overdispersion and unobserved heterogeneity, \( Y_{b,c,k} \) is modeled using a cluster-level beta-binomial likelihood,

\begin{equation}\label{eq:educ-betabinomial}
Y_{b,c,k}  ~|~  h_{b,c,k} \sim \mbox{BetaBinomial}\left(n_{b,c,k},h_{b,c,k},\phi \right)
\end{equation} 
where $\phi$ is the overdispersion parameter, parameterized so that
$$\mbox{Var}(Y_{b,c,k}) = h_{b,c,k}(1-h_{b,c,k})/\phi.$$
A detailed derivation and discussion of the beta-binomial likelihood is provided in Section \ref{sec:supp-beta-bin-derive} in the supplemental materials.

The logit hazard is then modeled using the following spatiotemporal model:

\begin{eqnarray}
\log \left( \frac{h_{b,c,k}}{1-h_{b,c,k}} \right)&=& 
\underbrace{\beta_k}_{\substack{
\text{Grade-specific}\\
\text{Intercept}
}} +
\underbrace{\phi_b + \nu_b}_{\substack{
\text{Birth Cohorts}\\
\text{Main Effects}
}}+ \underbrace{u_i }_{\substack{
\text{Spatial Effect} 
}} +
\underbrace{\zeta_{i,b} }_{\substack{
\text{Space-Birth Cohorts}\\
\text{Interaction} 
}} 
\label{eq:educ-cluster-model-mean-admin2}
\end{eqnarray}

\noindent with the model components defined as follows:

\begin{itemize}
    \item \textbf{Temporal Main Effects for Birth Cohorts:}
    
    where \( \phi_b \) and \( \nu_b \) are the structured (first-order random walk, RW1) and unstructured (independent and identically distributed, IID) temporal effects for birth cohorts, respectively. 
    
    \item  \textbf{Spatial Effects for Areas:}

    The term \( u_i \) represents the spatial random effect for area \( i \), where cluster \( c \) is located. We adopt the BYM2 model \citep{riebler2016an}, a reparametrization of the Besag-York-Mollie (BYM) model \citep{besag1991bayesian}, which decomposes \( u_i \) into an IID component \( e_i \) and a spatially structured component \( S_i \):  
\begin{equation*}
u_i = \sigma (\sqrt{1-\phi} e_i + \sqrt{\phi} S_i),
\end{equation*}
where \( S_i \) follows a scaled intrinsic conditional autoregressive (ICAR) prior. We use penalized complexity (PC) priors \citep{simpson2017penalising} for the hyperparameters $\sigma$ (total standard deviation) and $\phi$ (the proportion of the variation that is spatial).

 \item  \textbf{Space-Time Interaction:}

The interaction term \( \zeta_{i,b} \) accounts for deviations from the main effects of birth cohorts and spatial components and is modeled as a type-IV space-time interaction \citep{knorr2000bayesian}. 
\end{itemize}

Posterior inference is conducted using the Integrated Nested Laplace Approximation (INLA) method, implemented in the R package \texttt{INLA} \citep{rue2009approximate}.

As with the survey GLM model, Equation \eqref{eq:educ-cluster-model-mean-admin2} assumes that the impact of grade-level is consistent across all birth cohort-area combinations (i.e., it too makes a proportional odds assumption). 

\subsubsection{Urban/Rural Stratification}

The spatial model specified in Equation (\ref{eq:educ-betabinomial}) does not incorporate survey design weights, and thus cannot fully account for the sampling design. Ignoring urban/rural stratification leads to significant biases \citep{wu2024modelling,dong2021modeling}, which arise from the combination of two factors: the oversampling of urban clusters, a common feature of DHS surveys, and the pronounced differences in patterns between urban and rural areas. In the context of education, this issue is especially critical given the substantial urban/rural disparities in educational attainment, particularly in LMICs \citep{UNESCO2024}.

To address this, we construct separate beta-binomial models for urban and rural clusters at the domain level, ensuring that distinct urban- and rural-specific hazards, \( h_{b,c,k}^{U} \) and \( h_{b,c,k}^{R} \), are estimated based on Equations \eqref{eq:educ-betabinomial} and \eqref{eq:educ-cluster-model-mean-admin2}. For simplicity, we collapse the area and birth cohort indices into a single domain index \( d \). These hazard estimates are then used to derive separate urban and rural expected years of schooling (UYS) estimates, denoted as \( \mu_{d}^{U} \) and \( \mu_{d}^{R} \). 

Since the models are fit separately for urban and rural clusters, the resulting estimates must be combined to obtain an overall domain-level estimate. This requires an \textbf{aggregation step}:
\begin{equation*}
\mu_d = r_d \times \mu_{d}^{U} + (1 - r_d) \times \mu_{d}^{R},
\end{equation*}  
where \( r_d \) represents the proportion of the urban population in domain \( d \). 

We obtain \( r_d \) based on a recent approach that employs classification models to reconstruct urban/rural partitions \citep{wu2024modelling}. We provide further details on the methodology for estimating these proportions in Section \ref{sec:supp-UR-stratification} in the supplemental materials.

\section{Simulation}
\label{sec:sims}
\subsection{Setup}

To evaluate the impact of right-censoring on birth cohort-specific estimates of ultimate years of schooling (UYS), we conducted a simulation study designed to reflect the patterns of incomplete data observed in the 2022 Tanzania DHS data. Unlike traditional simulation studies, which involve generating data by assuming specific parameters and testing whether methods recover these parameters, our approach aims to replicate a real-world scenario.. The primary objective is to quantify and mitigate bias introduced by right-censoring in a realistic setting. Based on the empirical distribution of observed data, we construct censored scenarios and evaluate whether our proposed statistical methods can mimic estimators under full data availability. 

TheAn simulation study is based on the 2022 Tanzania DHS survey data for females, focusing on four consecutive 5-year birth cohorts: 25--29, 30--34, 35--39 and 40--44 years, measured at the time of the survey. For these age groups, naive weighted estimation of UYS is considered reliable since all individuals in these cohorts have completed their schooling by the time of the survey, leading to effectively no right-censoring. Thus, naive weighted estimates obtained from such complete data conditions serve as the ground truth for comparison.

To simulate right-censoring, we imposed conditions similar to those faced by the cohort aged 15--19 at the time of the survey. We define \( T_{\text{survey}}^* \) as the adjusted survey time in the simulation. For example, for the 25--29 cohort, we simulated data collection 10 years earlier, shifting \( T_{\text{survey}}^* \) to when this cohort was still within the schooling age range. Similarly, for other cohorts, data collection was conceptually shifted to ages 15–19, introducing right-censoring. Importantly, in each simulation scenario, only one age group was artificially censored, while the remainder of the data remained unmodified to ensure the robustness of comparison across scenarios.

For each individual, school exit time (discontinuation/completion) was determined as:
\begin{equation}
\label{eq:cal-exit-time}
T_{\text{exit}} = T_{\text{birth}} + E + T,
\end{equation}
where \( T_{\text{birth}} \) is the birth year, \( E \) is the school entrance age, and \( T \) is UYS (fully observed for individuals aged $\ge$25 in Tanzania 2022 DHS). Right-censoring was imposed by comparing \( T_{\text{exit}} \) with the adjusted survey time \( T_{\text{survey}}^* \), classifying an individual as censored ($\delta^*=0$) if:
\begin{equation*}
    T_{\text{exit}} > T_{\text{survey  }}^* .
\end{equation*}
This setup mirrors real-world scenarios where younger individuals in a DHS survey may still be in school at the time of data collection. While \( T_{\text{birth}} \), \( T \), and \( T_{\text{survey}}^* \) are directly recorded in the survey, \( E \) is unobserved. To generate realistic censoring patterns, we explicitly modeled variations in school entrance age, in particular delayed school entry, a key driver of right-censoring in UYS. The only source of randomness in our simulation is \( E \), which was sampled from a modeled distribution of school entrance ages. Further details on the delayed school entry model and other technical aspects of the simulation setup are provided in Section \ref{sec:supp-sim-setup} in the supplemental materials. 

Each simulation scenario was repeated 10 times, with results averaged to obtain stable estimates. This approach allowed us to systematically evaluate the bias introduced by end-of-study censoring and compare the performance of weighted estimation approaches against our proposed discrete-time survival methods for handling censored data.

\subsection{Results}

\begin{figure} [!ht]
    \centering
    \includegraphics[clip, trim=0cm 0.3cm 0cm 0.2cm, width=1\linewidth]{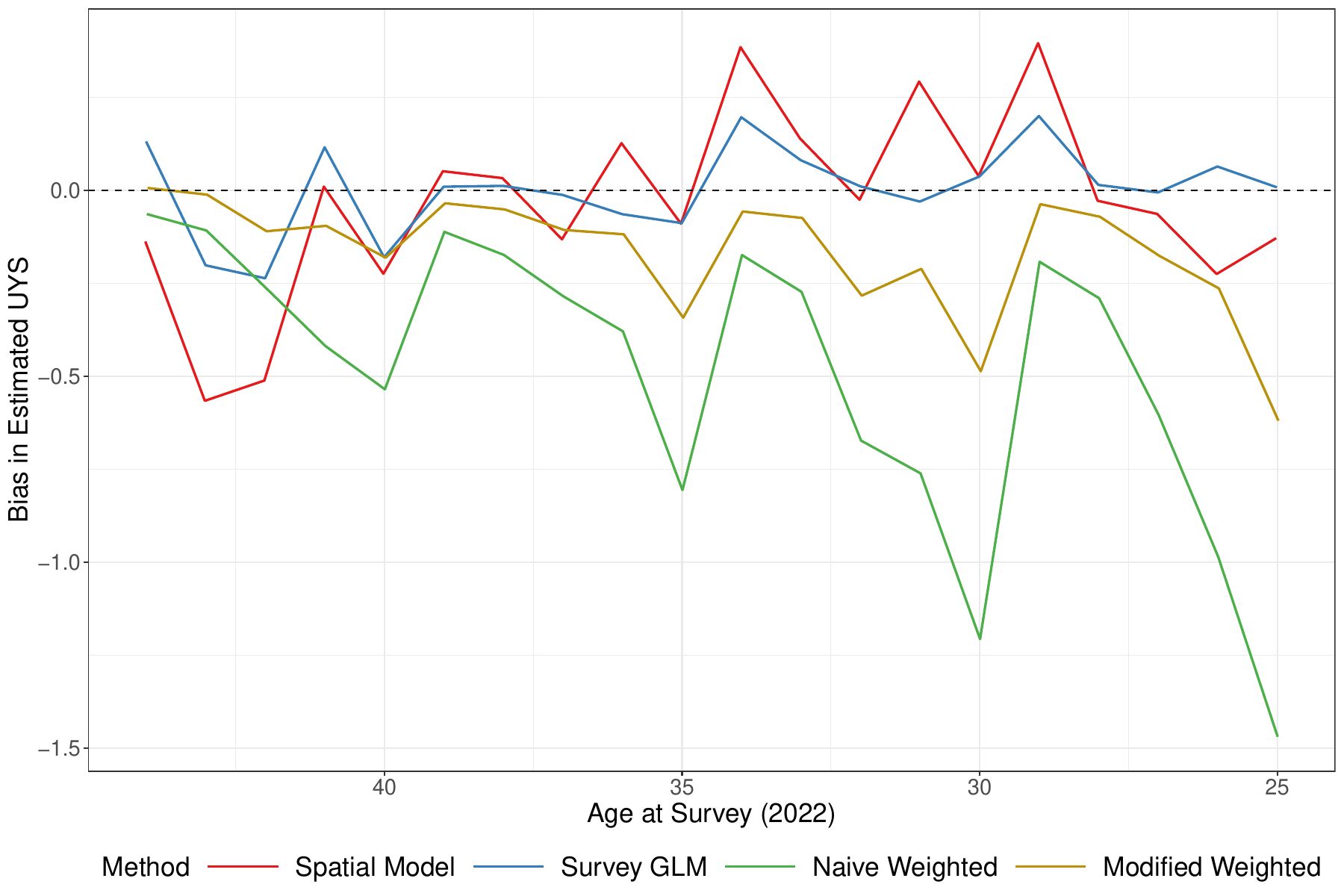}
    \caption{Bias in UYS under right-censoring across different estimation methods. The $x$-axis represents age at the time of the 2022 Tanzania DHS, with younger ages on the right and older ages on the left to align with birth cohorts. The horizontal dashed line at zero indicates an unbiased estimate relative to naive weighted estimates from the full data.}
    \label{fig:sim-res-yearly}
\end{figure}

Table \ref{tab:sim_bias_summary_TZA} and Figure \ref{fig:sim-res-yearly} presents the bias estimates from the simulation study, assessing the impact of right-censoring on UYS estimates for females in the 2022 Tanzania DHS. The results compare four statistical approaches: (a) \textbf{naive weighted estimation}, which uses survey-weighted estimates without any adjustments for right-censoring (Section \ref{sec:direct-unadj}); (b) \textbf{modified weighted estimation}, a partially adjusted direct estimation approach (Section \ref{sec:direct-modified}); (c) \textbf{survey GLM}, a survey-weighted generalized linear model designed to incorporate right-censoring adjustments (Section \ref{sec:survey-GLM}); and (d) \textbf{spatial model} (Section \ref{sec:unit-level-model}), implemented at the Admin-1 level, with estimates aggregated to the national level based on the relative population distribution across regions.

For each method, UYS estimates were computed at the national level, stratified by yearly birth cohort, which corresponds to age at the time of the 2022 survey. Although right-censoring was introduced at the 5-year cohort level (e.g., 25--29, 30--34), individuals within each cohort were affected differently. For example, in the 25--29 age group, right-censoring was simulated as if the survey had taken place 10 years earlier in 2012. Within this group, those aged 25 in 2022 were 15 in 2012, making them more likely to be in school and thus censored, whereas those aged 29 in 2022 were 19 in 2012, facing a lower risk of censorship. To capture these differences, UYS was calculated separately for each birth year (or equivalently, age at the 2022 survey). 

Bias was defined as the difference between the UYS estimate from the censored dataset and the corresponding naive weighted estimate from the original dataset, which serves as the ground truth.

\begin{table}[!ht]
\centering
\caption{Summary for bias in UYS estimates under right-censoring across estimation methods in the 2022 Tanzania DHS. The table reports overall and stratified bias by age at censoring (simulation-imposed) and age group at the 2022 survey. Negative values indicate underestimation.}

\begin{tabular}{lrrrr}
\hline
& \textbf{Naive} & \textbf{Modified} & \textbf{Survey} & \textbf{Spatial} \\
 & \textbf{Weighted} & \textbf{Weighted} & \textbf{GLM} & \textbf{Model} \\
\hline
Average Bias & -0.49 & -0.17 & 0.01 & -0.03 \\
\cmidrule(lr){1-5}
\textbf{Age at Censoring} &  & & &  \\
\hline
15 & -1.00 & -0.41 & -0.06 & -0.10 \\
16 & -0.64 & -0.17 & 0.02 & 0.06 \\
17 & -0.46 & -0.17 & -0.06 & -0.18 \\
18 & -0.21 & -0.05 & -0.02 & -0.10 \\
19 & -0.14 & -0.03 & 0.13 & 0.17 \\
\cmidrule(lr){1-5}
\textbf{Age Group} &  & & &  \\
\hline
25--29 & -0.71 & -0.23 & 0.06 & -0.01 \\
30--34 & -0.62 & -0.22 & 0.06 & 0.17 \\
35--39 & -0.35 & -0.13 & -0.03 & -0.01 \\
40--44 & -0.28 & -0.08 & -0.07 & -0.29 \\
\hline
\end{tabular}
\label{tab:sim_bias_summary_TZA}
\end{table}

The results indicate that naive weighted estimation leads to systematic underestimation, with an average bias of -0.49 years, underscoring the strong impact of right-censoring. The modified weighted estimation reduces this bias to -0.17, demonstrating partial improvement, but fails to fully correct for censoring due to incomplete educational trajectories. In contrast, the discrete-time survival models, survey GLM (0.01) and the spatial model (-0.03), effectively mitigate bias, producing the most accurate estimates.

Bias is most pronounced for individuals aged 25, 30, 35, and 40 at the time of the 2022 survey, as they were subject to right-censoring at age 15 under the artificially imposed survey timing. At this age, a large proportion of individuals were still in school, leading to severe underestimation of UYS --- up to \textbf{one full year} in the naive weighted estimates. Given that the national UYS estimates in Tanzania range between 5 to 9 years (as shown in Figure \ref{fig:natl-overall-educ-yrs-TZA-female}), this represents a substantially large bias. In contrast, the modified weighted estimation alleviates some of this bias, while survey GLM and the spatial model effectively resolve it.

Bias trends are similar across age groups (relative to the original 2022 Tanzania DHS), with younger cohorts experiencing greater underestimation due to higher likelihood of being censored. The 25--29 cohort exhibits the largest bias under naive weighted estimation (-0.71), reinforcing that higher UYS increase susceptibility to right-censoring. Again, survey GLM (0.06) and the spatial model (-0.01) substantially reduce this bias, confirming their superior performance in addressing censoring effects. In some cohorts, the spatial model shows slightly larger deviations from the full data estimates (e.g., 30--34: 0.17, 40--44: -0.29). This is likely due to the model’s smoothing of temporal trends, which pulls estimates toward the overall trend and reduces local fluctuations. For example, the bump observed around the 1978--1979 birth cohort (within the 40--44 age group) in the full data is smoothed out in the spatial model, as shown in Figure \ref{fig:natl-overall-educ-yrs-TZA-female}. A similar effect is expected in the simulation, contributing to the observed bias in this age group. However, this bias is not systematic, and the overall low bias across age groups confirms the robustness and reliability of the spatial model.

These findings highlight the clear limitations of weighted estimation for UYS in the presence of right-censoring, particularly for younger cohorts and/or cohorts with higher education attainment. While partial adjustments reduce bias, methods that explicitly model the full educational trajectory, such as survey GLM and the spatial model, yield more accurate estimates.

\section{Data Application: Female Education in Tanzania}
\label{sec:results}
This section presents the findings from our application of the proposed methods to UYS for females in Tanzania. We focus on the female population as female educational attainment is not only a key indicator in education monitoring, but is also routinely used in the study of fertility, child mortality, and other development outcomes (see for example, \citealt{fuchs2010education}). Of course, our methods can be applied to the male population, other subpopulations, or the overall population depending on the goals of an analysis.

The results are organized into two parts. First, we analyze national-level estimates separately for each DHS survey conducted after 2000. This comparison allows us to assess the effects of right-censoring and evaluate the performance of our methods by generating UYS estimates from earlier surveys and validating them against more recent data, since censored cohorts in older surveys mature into completed cohorts in recent surveys.

The second part provides a detailed analysis based on the most recent DHS 2022 survey, where we assess UYS estimates across multiple spatial and temporal scales. Specifically, we investigate regional and urban-rural disparities, analyze birth cohort trends, and compare estimates derived from different modeling approaches.

\subsection{National-Level Estimates for UYS Across Surveys}

We apply our proposed methods to each Demographic and Health Survey (DHS) conducted in Tanzania since 2000, specifically DHS 2004, DHS 2010, DHS 2015, and DHS 2022. To maintain clarity, we present results using two estimation methods: \textbf{naive weighted estimates}, which are biased due to right-censoring, and the \textbf{survey GLM}, which correctly accounts for censoring.  Given the large sample sizes at the national level, we adopt the survey GLM method, reserving spatial models for more granular subnational analyses.

Given the birth-cohort-specific nature of UYS, we use historical surveys to validate our estimates. For a given birth cohort, we compare UYS estimates across multiple surveys: younger cohorts in earlier surveys will age through the educational system and present with less or no censoring in later surveys. As a concrete example, individuals from the 1995 birth cohort were 15 years old in DHS 2010 and 27 years old in DHS 2022. While UYS estimates for this birth cohort based on DHS 2010 are subject to right-censoring, those based on DHS 2022 can serve as a fully observed benchmark, allowing us to evaluate the accuracy of earlier projections using our discrete-time survival models. This approach mirrors our simulation study, where artificially censored cohorts were later validated using fully observed data.

\begin{figure} [!ht]
    \centering
    \includegraphics[clip, trim=0cm 0cm 0cm 0cm, width=0.98\linewidth]{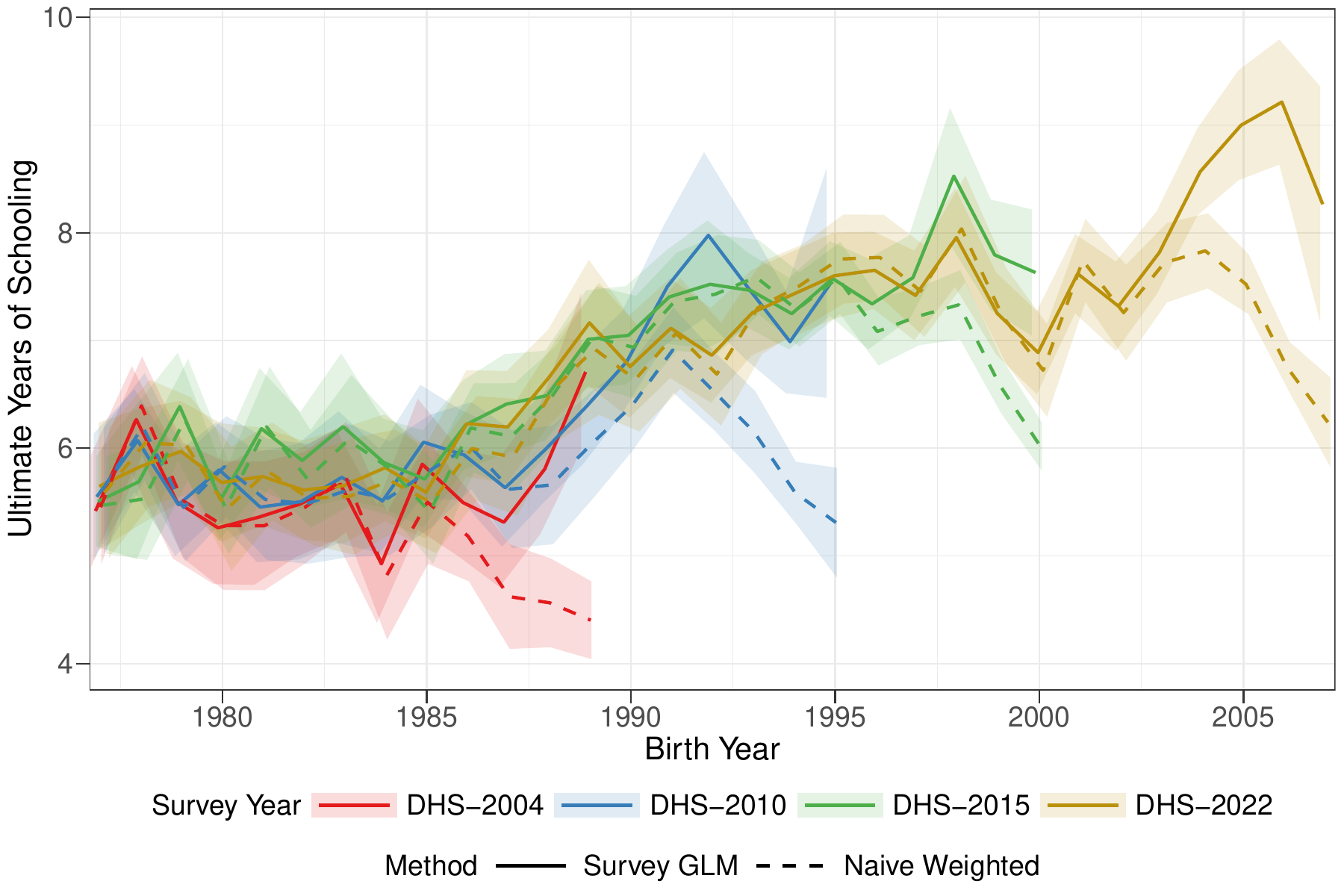}
    \caption{National-level birth-cohort specific UYS estimates across DHS surveys in Tanzania (DHS 2004, 2010, 2015 and 2022), using survey GLM (solid lines) and naive weighted estimation (dashed lines). Differences in UYS trends across surveys reflect variations in estimation methods and the effects of right-censoring.}
    \label{fig:natl-UYS-multiple-surveys}
\end{figure}

Figure \ref{fig:natl-UYS-multiple-surveys} presents national-level UYS estimates by birth cohort, derived separately from each DHS survey. The results reveal a consistent pattern, where naive weighted estimates (dashed lines) systematically underestimate UYS for individuals aged 15--19 at the time of each survey. This underestimation, particularly evident in the sharp drop at the right end of each dashed line, is a direct consequence of right-censoring, as these individuals had not yet completed their education when surveyed.

In contrast, survey GLM estimates (solid lines) are more consistent across surveys. The estimates based on earlier surveys closely align with estimates from later surveys, where schooling for these cohorts is fully observed. Despite survey-specific fluctuations, the survey GLM method effectively captures the overall trend in UYS across birth cohorts. For instance, the sharp rise in UYS among the 1985–1989 birth cohorts is well reflected in the survey GLM estimates, whereas naive weighted estimation severely underestimates this increase. These findings highlight the importance of properly accounting for right-censoring, as failure to do so will result in misleading trends in UYS estimates for younger cohorts.

\subsection{Detailed Results for 2022 Tanzania DHS}

In this section, we analyze the most recent DHS 2022 survey, which provides the latest estimates of UYS in Tanzania. We examine education attainment trends over birth cohorts, assess regional and urban-rural disparities, and compare estimates across different modeling approaches. Additionally, we present additional statistical metrics naturally derived from our modeling process, including educational attainment distributions, which offering further insight into education patterns in Tanzania.

\subsubsection{National UYS Trend across Birth Cohorts}

We begin by revisiting Figure \ref{fig:natl-overall-educ-yrs-TZA-female}, which presents national-level UYS estimates for female by yearly birth cohort across different estimation methods: naive \textbf{weighted estimation}, \textbf{modified weighted estimation}, \textbf{survey GLM}, and \textbf{spatial model}. This comparison highlights how each method addresses right-censoring for the most recent birth cohorts, where incomplete schooling data can introduce bias. Particularly for birth cohort 2003-2007, naive weigthed estimation method systematically underestimates UYS. Modified weighted estimation provides some correction but still exhibits noticeable discrepancies. In contrast, the discrete-time survival models, namely survey GLM and the spatial model yield more consistent and realistic trends, aligning well with expected educational trajectories.

All methods capture the increasing trend in UYS for cohorts born between 1975 and 2000, reflecting long-term improvements in educational attainment, with UYS rising by about three years over this 25-year period. Notably, the spatial model imposes a smoothing structure across yearly birth cohorts, producing a more stable trend that is less sensitive to random temporal fluctuations. This smoothing makes the spatial model more preferable in retrieving broader trends in educational attainment over birth cohorts without being overly influenced by year-to-year variability.

\subsubsection{Subnational UYS Estimates across Birth Cohorts}

We extend our analysis to subnational UYS estimation at both Admin-1 (regional) and Admin-2 (district) levels in Tanzania, which consists of 31 regions and 186 districts. We adopt GADM database \citep{gadm} to determine geographical boundaries and cluster assignments.

For subnational analysis, we focus on results produced by the spatial model defined in Equation (\ref{eq:educ-cluster-model-mean-admin2}), as it offers a more flexible structure to account for region $\times$ birth cohort interactions---an aspect the survey GLM struggles to accommodate because of data sparsity. A key advantage of the spatial model is its ability to address data sparsity, which is particularly relevant given that about 10\% of districts have no sampled observations at all, and even in surveyed regions, data are not always available for every birth cohort. These gaps pose identifiability challenges for the survey GLM, whereas the spatial model leverages smoothing structures and captures complex interactions between birth cohorts and geographic units, stabilizing estimates in data-sparse areas while preserving local variations.

\begin{figure} [!ht]
    \centering
    \includegraphics[clip, trim=0cm 0cm 0cm 0cm, width=0.98\linewidth]{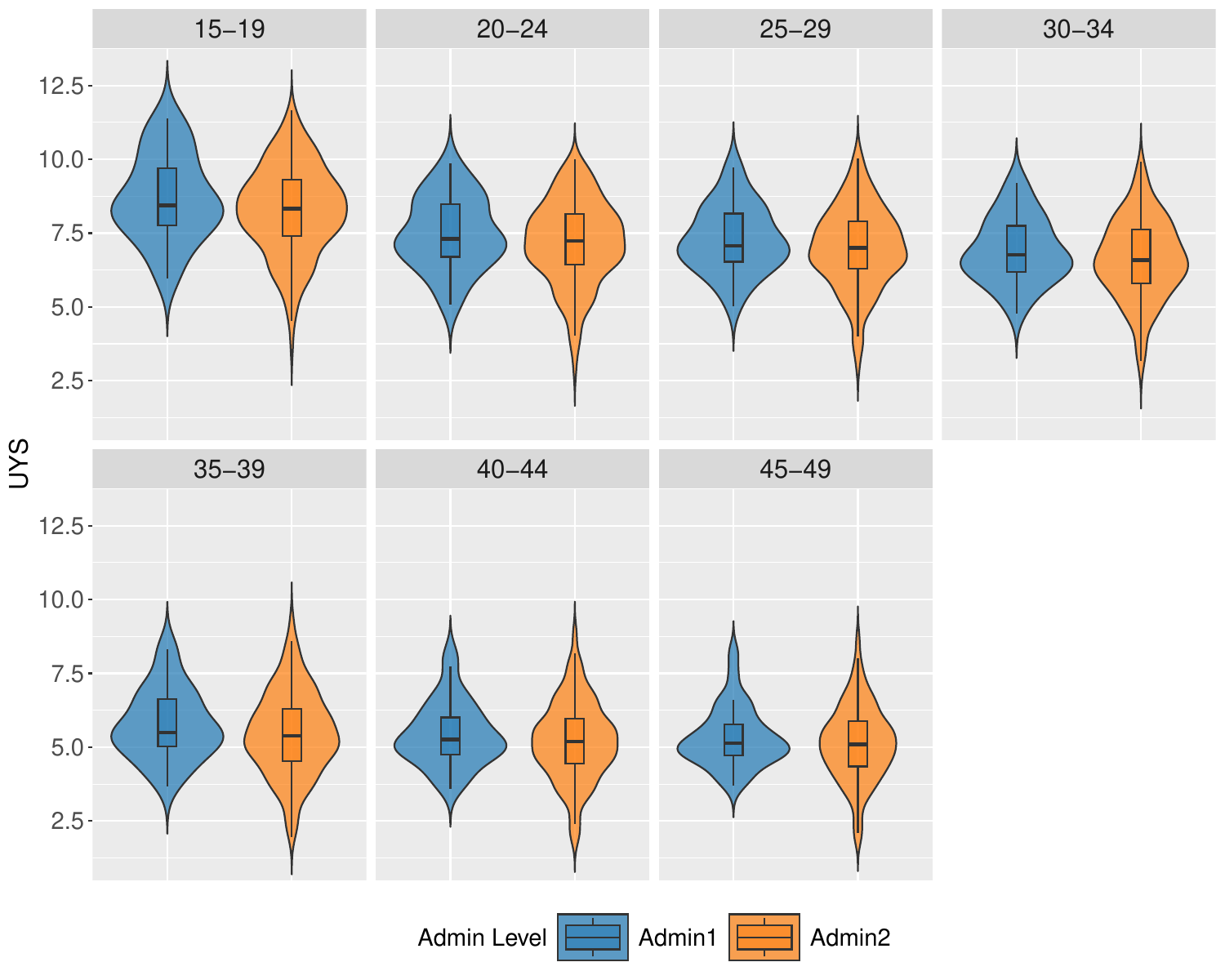}
    \caption{Subnational variation in UYS across 5-year age groups.}
    \label{fig:TZA-2022-subnatl-box-violin}
\end{figure}

Figure \ref{fig:TZA-2022-subnatl-box-violin} presents the subnational distribution of estimated UYS across different age groups at both the Admin-1 and Admin-2 levels using a combination of box and violin plots. These visualizations capture the range of subnational variation, highlighting disparities in educational attainment within each cohort.

A clear trend emerges across age groups: UYS decreases with age, indicating that younger cohorts (e.g., 15--19 years) are expected to attain higher levels of education than older cohorts (e.g., 45--49 years). These cohort differences provide strong evidence of progressive improvements in educational attainment over generations, aligning with trends observed in national-level estimates. 

\begin{figure} [!ht]
    \centering
    \includegraphics[clip, trim=0cm 2.5cm 0cm 2cm, width=0.98\linewidth]{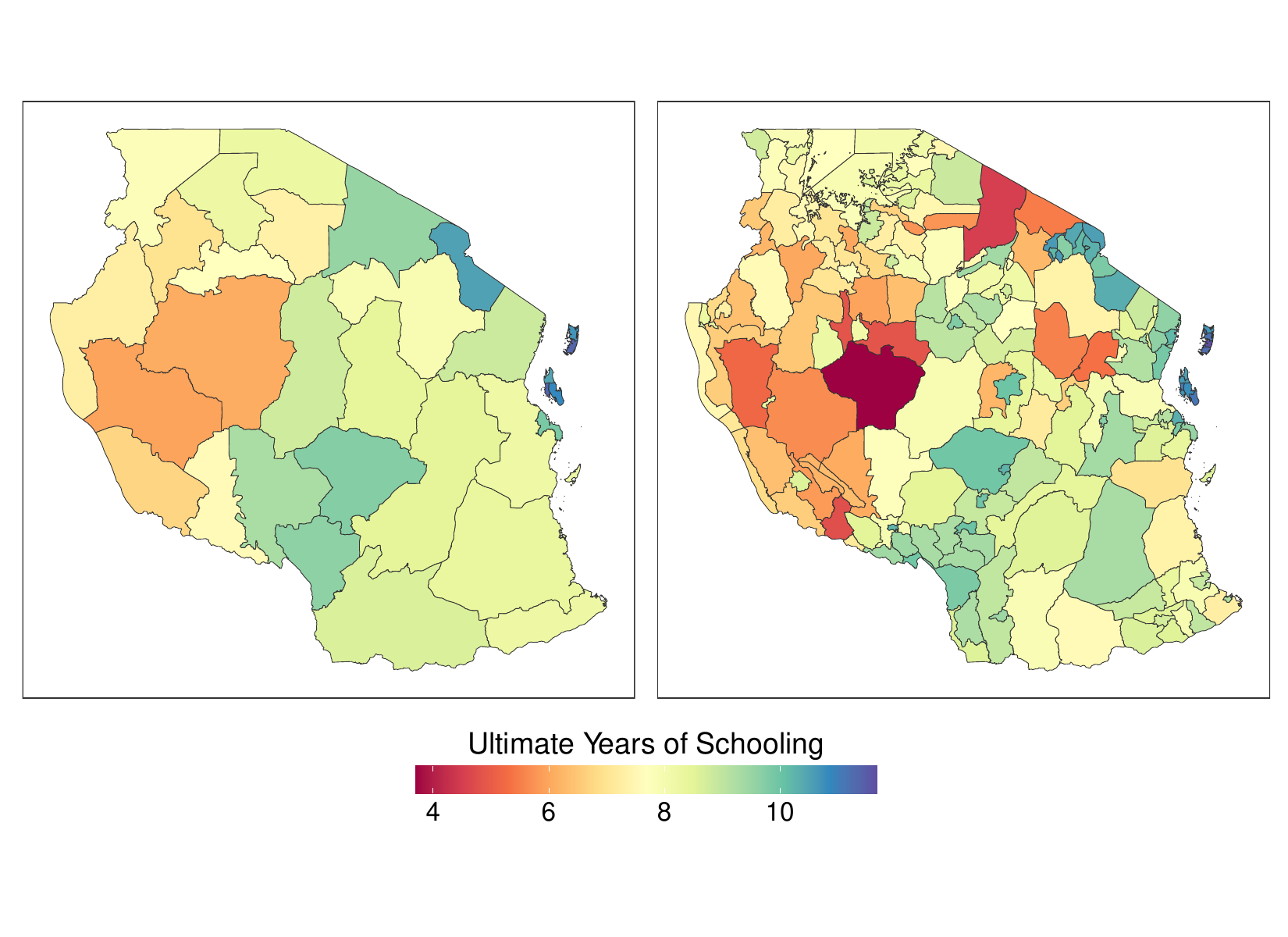}
    \caption{Comparison of Admin-1 and Admin-2 estimates of UYS for the most recent birth cohort (ages 15--19) in the 2022 Tanzania DHS. }
    \label{fig:admin1-admin2-5yr-UYS-15-19}
\end{figure}

Notably, subnational disparities in education are more pronounced at the Admin-2 level than at the Admin-1 level, particularly at the lower end of the distribution, where regions with low educational attainment diverge more sharply. Figure \ref{fig:admin1-admin2-5yr-UYS-15-19} comparing Admin-1 and Admin-2 estimates for UYS in the most recent birth cohort (ages 15--19 in the 2022 survey) visually illustrates this pattern. Further, the map suggests that finer geographic units reveal deeper inequalities that may be masked at coarser levels of aggregation. The persistence of substantial regional gaps across all age groups underscores the importance of targeted interventions to address localized barriers to education. 

In addition, although an overall increasing trend in educational attainment is evident across birth cohorts, the magnitude of progress varies considerably across regions. To further examine these discrepancies, we provide additional visualizations in Section \ref{sec:supp-figure-subnational-UYS} and \ref{sec:supp-figure-extra-metric} of the supplemental materials.

\subsubsection{Urban-Rural Disparities}

An important aspect of our analysis is the examination of urban-rural differences in UYS estimates. This distinction arises naturally as an intermediate product of our modeling framework for spatial models, where we account for the survey design by fitting separate models for urban and rural areas before aggregating to overall estimates. As a result, we inherently expect significantly different urban-rural patterns, both in terms of the educational trajectories individuals follow through the school system and the overall generational improvements in education levels.

While we anticipate increases in UYS for both urban and rural populations over time, the rate of growth may differ, potentially leading to a widening or narrowing of the urban-rural gap. If urban education expands at a faster rate, the gap is likely to grow; conversely, if rural areas experience accelerated improvements, the disparity may diminish. In the case of Tanzania, recent policy initiatives, such as the Education Sector Development Plan (ESDP) proposed by the \cite{TanzaniaESDP2020}, have prioritized equitable access to quality education, aiming to reduce urban–rural disparities through expanded schooling opportunities and targeted interventions.  

\begin{figure} [!ht]
    \centering
    \includegraphics[clip, trim=0cm 1.5cm 0cm 1cm, width=0.98\linewidth]{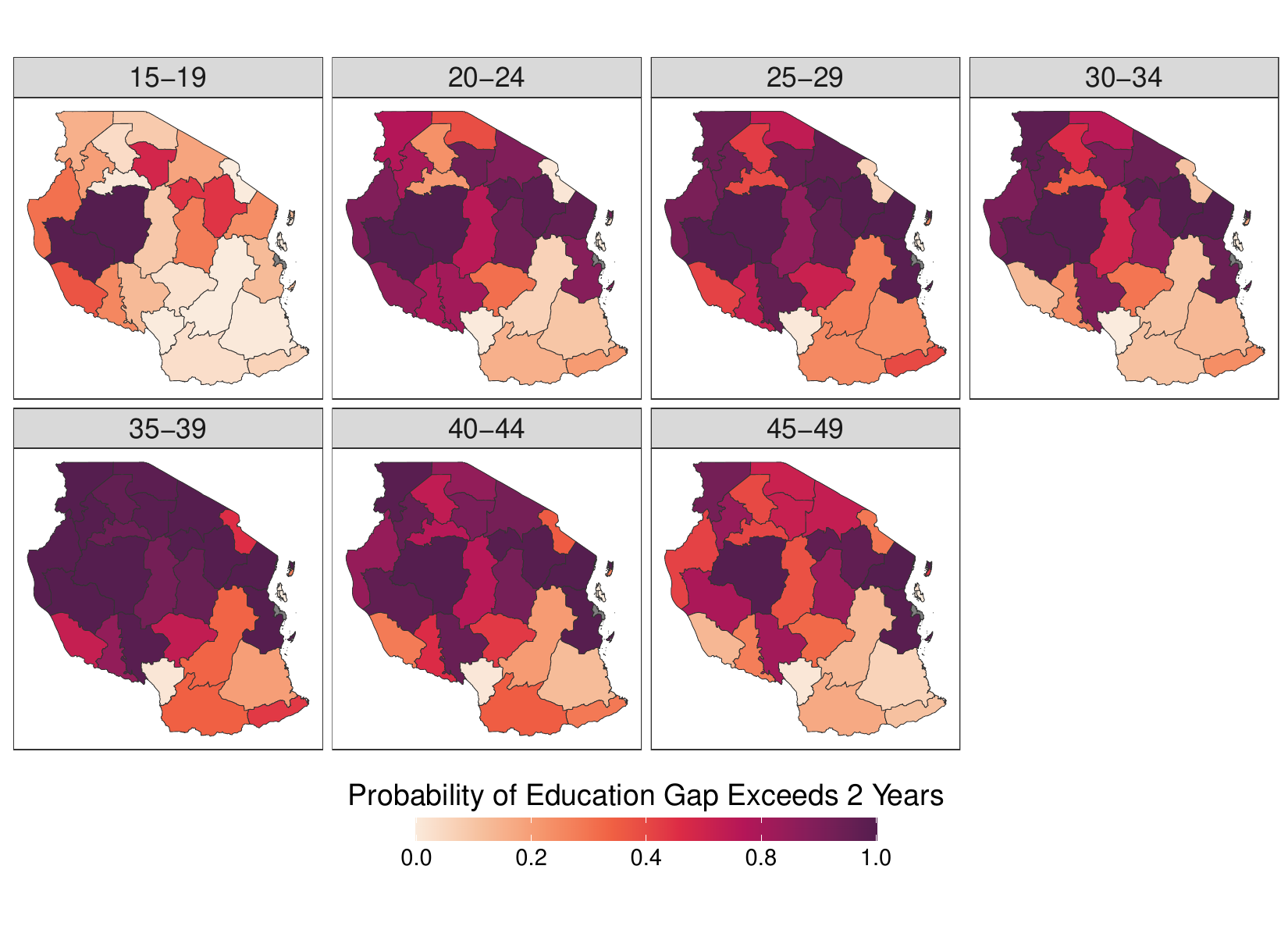}
    \caption{Probability that the urban–rural difference in UYS exceeds two years across regions and age groups. Dar es Salaam is excluded from the analysis as it is entirely urban and does not have rural estimates.}
    \label{fig:admin1-UR-gap-exceedance}
\end{figure}

To assess urban–rural disparities in ultimate years of schooling (UYS) based on our modeling approaches, we visualize the probability that the urban-rural difference in UYS exceeds two years across Admin-1 regions and age groups. These probabilities are derived from summarizing the posterior draws of our urban and rural estimates obtained through spatial models, inherently accounting for estimation uncertainties. A two-year difference in education represents a substantial gap, as it aligns with the duration of key educational stages in Tanzania, such as upper secondary school, and can reflect disparities in both progression through and completion of these critical levels.

The maps shown in Figure \ref{fig:admin1-UR-gap-exceedance} reveal notable regional variations in the probability that urban–rural differences in UYS exceed two years. For younger generations (e.g., 15--19 age group in 2022), the probabilities are generally lower, suggesting that recent cohorts have experienced more equitable educational opportunities between urban and rural areas. This trend aligns with Tanzania's policy initiatives aimed at reducing educational disparities. Historically, older age groups exhibit higher probabilities of exceeding the two-year threshold, indicating that inequalities were more profound and persistent in the past. Notably, regions such as Katavi and Tabora (central-western Tanzania) emerge as hotspots with significant urban–rural disparities in education, even for the most recent birth cohorts (ages 15--19 in 2022). These findings underscore the progress made in recent years while highlighting the need for continued efforts to address persistent socioeconomic disparities in educational attainment, in particular areas.

\begin{figure} [!ht]
    \centering
    \includegraphics[clip, trim=0cm 0cm 0cm 0cm, width=1\linewidth]{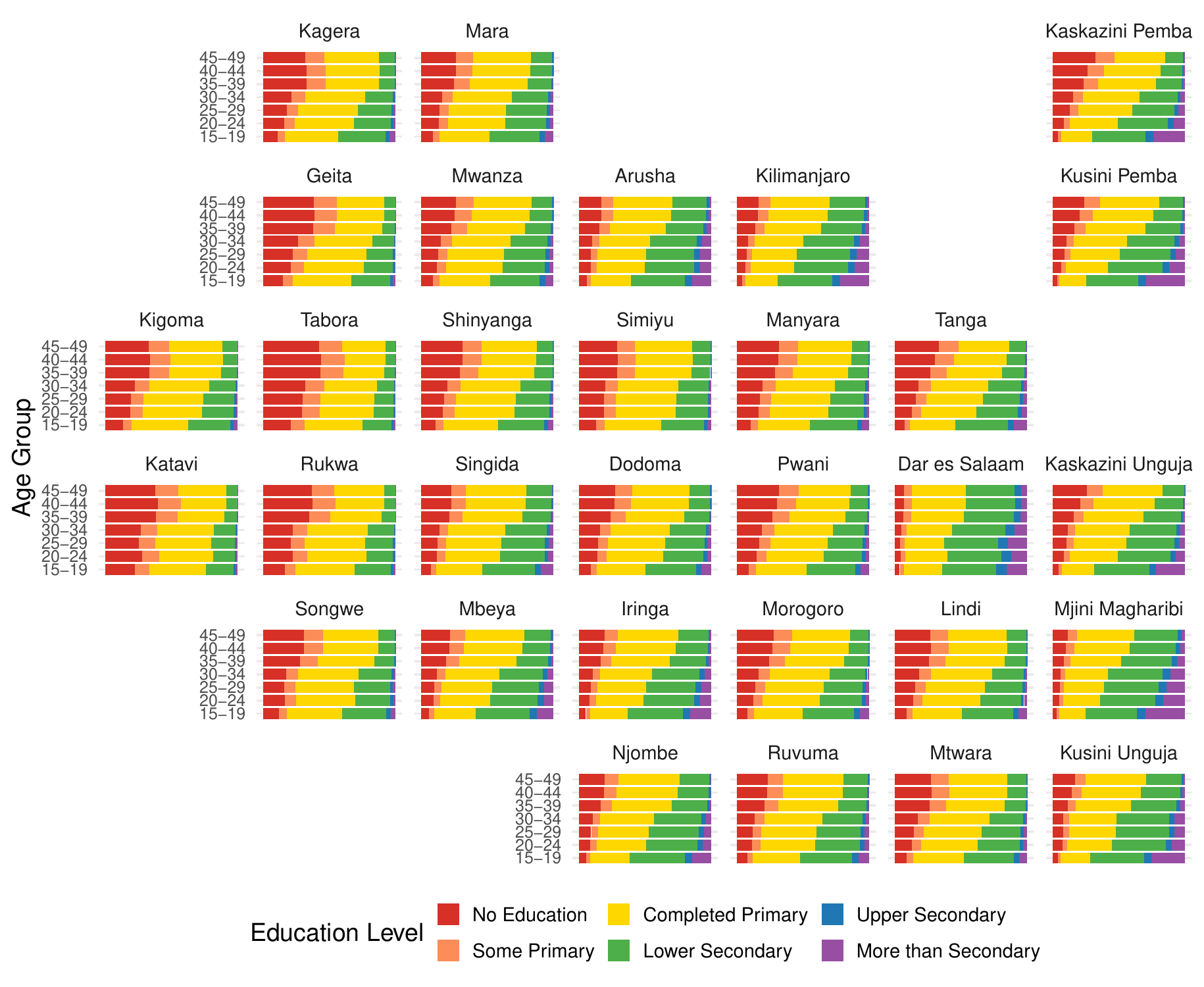}
    \caption{Distribution of education level by age group across Admin-1 regions. The panels are arranged to reflect the approximate geographic layout of Tanzania. }
    \label{fig:admin1-overall-educ-level-dist-geofacet}
\end{figure}

\subsubsection{Education Level Distribution}

A key strength of our methodological framework is its ability to go beyond the primary endpoint UYS. While UYS serves as a useful summary measure, our discrete-time survival model estimates piecewise continuation probabilities and the discrete distribution of grade attainment, offering a more granular view of school progression. This framework aligns with standard survival analysis concepts---marginal, conditional, and cumulative survival probabilities—all of which have meaningful interpretations in the context of education modeling.

Figure \ref{fig:admin1-overall-educ-level-dist-geofacet} presents the distribution of educational attainment across Admin-1 regions and age groups, grouping education levels into key stages: no education, some primary, completed primary, lower secondary, upper secondary, and more than secondary. The results show steady progress across generations, with younger cohorts (ages 15--19 and 20--24) achieving or expected to achieve higher education levels than older generations. The proportion of individuals with no formal education has declined sharply, particularly in island regions (displayed in the right-most column of the figure). However, progress has been slower in western Tanzania, such as Katavi and Tabora, where our previous findings highlighted substantial urban–rural disparities, indicating persistent challenges in school retention and progression.

Regional variation in higher education is also evident. Urban and economically developed regions, such as Dar es Salaam, Kilimanjaro, and the islands, have been projected to have more females obtain education beyond secondary school. In contrast, many rural regions still show limited transitions beyond secondary schooling, emphasizing the need for targeted policies to improve access to higher education.

Beyond this visualization, our framework allows for additional statistical summaries, such as cumulative probabilities of attaining at least a given education level, providing deeper insights into school progression and completion. These supplementary analyses, presented in Section \ref{sec:supp-figure-extra-metric} in the supplemental materials, further quantify regional disparities and trends across birth cohorts in educational attainment for particular areas.

\section{Discussion}
\label{sec:discussion}

In this paper, we have introduced ultimate years of schooling (UYS), a novel metric for tracking educational attainment by capturing birth cohort-specific educational trajectories and intergenerational progress. We have developed a methodological framework for estimating UYS that employs a discrete-time survival model to correct for bias from right-censoring, a key limitation in weighted estimation using household surveys. By adopting techniques from SAE, we can naturally extend our approach for subnational estimation, enabling the identification of regional
disparities in educational attainment at fine geographic scales. Moreover, modeling UYS as a time-to-event outcome enables us to track educational trajectories with high granularity by capturing continuation probabilities and grade specific attainment distributions. This innovative method offers deeper insights into subnational and intergenerational school progression and retention patterns, positioning our framework as a powerful tool for advancing education research and policy design.

The methodological advancements in this study effectively correct for right-censoring bias. Our simulation study (Section \ref{sec:sims}) confirms that both naive and modified weighted estimation underestimates UYS for younger cohorts, while our discrete-time survival model recovers unbiased estimates. Empirical results from historical surveys in Tanzania (Figure \ref{fig:natl-UYS-multiple-surveys}) further validate our modeling framework, demonstrating that our modeled estimates align with fully observed data from subsequent surveys, whereas weighted estimation remains biased. In addition, our results highlight the critical need for subnational analysis in education research. National-level trends often mask substantial disparities, as demonstrated in our data application to the 2022 Tanzania DHS, where the results indicate significant variation in UYS across Admin-1/2 regions and urban-rural settings. By incorporating SAE techniques via a Bayesian spatiotemporal framework, our models reveal localized inequalities that are persistent across birth cohorts, reflecting the influence of local socioeconomic conditions on educational opportunities. The ability to estimate UYS at fine geographic scales is especially valuable in LMICs, where education policies and resource distribution vary widely, underscoring the importance of targeted interventions at the district level.

Our framework is highly flexible. As we model discontinuation odds directly, various measures of attainment (including EYS and MYS25) and attrition can be recovered by aggregating our model parameters appropriately. In addition to its use as a monitoring tool, our model can also be viewed as a vehicle for conducting inferential studies. By introducing covariates (at either the individual or areal levels) into Equation \eqref{eq:svy-weighted-GLM-link2} or Equation \eqref{eq:educ-cluster-model-mean-admin2}, education researchers can explore the impact of various contextual and policy factors on  discontinuation odds, and use such insights to drive data-informed policies.

We conclude with two limitations of our work. As with all modelling projects, our estimates can only be as good as the data we rely on. Household surveys can struggle to sample from important vulnerable populations such as street children and those living in slum communities \citep{thomson2014system}. On the presumption that attainment is materially lower in excluded populations, their omission may propagate as upwards biases in UYS estimates. A second limitation concerns the proportional odds assumption made in the discrete-time survival model. While this assumption does appear to be reasonable in the context of Tanzania (see Section \ref{sec:supp-prop-odds-assumption} of the supplemental materials), future research is needed to understand the contexts under which the assumption may be violated. For example, in crisis situations where there is significant displacement (such as due to conflict or famine) or school closures (such as due to the COVID-19 pandemic), discontinuation patterns may shift abruptly. By introducing temporal interactions or random effects to the grade-specific intercepts, our framework can accommodate departures from proportionality, though the data requirements for such an extension are significant and again warrant further study. 

\section*{Acknowledgments}

Jon Wakefield was supported by R37 AI029168 and R01 HD112421 from the National Institutes of Health of the United States. Ameer Dharamshi was supported by the Natural Sciences and Engineering Research Council of Canada. Yunhan Wu was supported by R37 AI029168 from the National Institutes of Health of the United States.

\bibliographystyle{natbib} 
\bibliography{refs}

\begin{thebibliography}{}

\bibitem[Allison(1982)Allison]{allison1982discrete}
Allison, P.~D. (1982).
\newblock Discrete-time methods for the analysis of event histories.
\newblock {\em Sociological methodology\/}, {\bf 13}, 61--98.

\bibitem[Barro and Lee(1993)Barro and Lee]{barro1993international}
Barro, R.~J. and Lee, J.-W. (1993).
\newblock International comparisons of educational attainment.
\newblock {\em Journal of Monetary Economics\/}, {\bf 32}, 363--394.

\bibitem[Besag {\em et~al.}(1991)Besag, York, and Mollié]{besag1991bayesian}
Besag, J., York, J., and Mollié, A. (1991).
\newblock Bayesian image restoration, with two applications in spatial statistics.
\newblock {\em Annals of the Institute of Statistical Mathematics\/}, {\bf 43}, 1--20.

\bibitem[Binder(1983)Binder]{binder1983on}
Binder, D.~A. (1983).
\newblock On the variances of asymptotically normal estimators from complex surveys.
\newblock {\em International Statistical Review / Revue Internationale de Statistique\/}, {\bf 51}, 279--292.

\bibitem[Brown(1975)Brown]{brown1975on}
Brown, C.~C. (1975).
\newblock On the use of indicator variables for studying the time-dependence of parameters in a response-time model.
\newblock {\em Biometrics\/}, {\bf 31}, 863--872.

\bibitem[Croft {\em et~al.}(2023)Croft, Allen, Zachary, {\em et~al.}]{croft2018guide}
Croft, T.~N., Allen, C.~K., Zachary, B.~W., {\em et~al.} (2023).
\newblock {\em Guide to {DHS} Statistics\/}.
\newblock ICF, Rockville, Maryland, USA.

\bibitem[Delprato and Frola(2022)Delprato and Frola]{delprato2022zones}
Delprato, M. and Frola, A. (2022).
\newblock Zones of educational exclusion of out-of-school youth.
\newblock {\em International Journal of Educational Development\/}, {\bf 88}, 102532.

\bibitem[Delprato {\em et~al.}(2024)Delprato, Chudgar, and Frola]{delprato2024spatial}
Delprato, M., Chudgar, A., and Frola, A. (2024).
\newblock Spatial education inequality for attainment indicators in sub-saharan {Africa} and spillovers effects.
\newblock {\em World Development\/}, {\bf 176}, 106522.

\bibitem[Dharamshi {\em et~al.}(2022)Dharamshi, Barakat, Alkema, and Antoninis]{dharamshi2022bayesian}
Dharamshi, A., Barakat, B., Alkema, L., and Antoninis, M. (2022).
\newblock A bayesian model for estimating sustainable development goal indicator 4.1.2: School completion rates.
\newblock {\em Journal of the Royal Statistical Society Series C: Applied Statistics\/}, {\bf 71}, 1822--1864.

\bibitem[Dong and Wakefield(2021)Dong and Wakefield]{dong2021modeling}
Dong, T.~Q. and Wakefield, J. (2021).
\newblock Modeling and presentation of vaccination coverage estimates using data from household surveys.
\newblock {\em Vaccine\/}, {\bf 39}, 2584--2594.

\bibitem[Fuchs {\em et~al.}(2010)Fuchs, Pamuk, and Lutz]{fuchs2010education}
Fuchs, R., Pamuk, E., and Lutz, W. (2010).
\newblock Education or wealth: Which matters more for reducing child mortality in developing countries?
\newblock {\em Vienna Yearbook of Population Research\/}, {\bf 8}, 175--199.

\bibitem[{Global Administrative Areas}(2022){Global Administrative Areas}]{gadm}
{Global Administrative Areas} (2022).
\newblock {GADM} database of global administrative areas, version 4.1.

\bibitem[Graetz {\em et~al.}(2018)Graetz, Zimmerman, Burstein, Biehl, Shields, Mosser, Casey, Deshpande, Earl, Reiner, Ray, Fullman, Levine, Stubbs, Mayala, Longbottom, Browne, Bhatt, and Hay]{graetz2018mapping}
Graetz, N., Zimmerman, A., Burstein, R., Biehl, M., Shields, C., Mosser, J., Casey, D., Deshpande, A., Earl, L., Reiner, R., Ray, S., Fullman, N., Levine, A., Stubbs, R., Mayala, B., Longbottom, J., Browne, A., Bhatt, S., and Hay, S. (2018).
\newblock Mapping local variation in educational attainment across {Africa}.
\newblock {\em Nature\/}, {\bf 555}, 48--53.

\bibitem[H{\'a}jek(1971)H{\'a}jek]{hajek1971discussion}
H{\'a}jek, J. (1971).
\newblock Discussion of ‘an essay on the logical foundations of survey sampling, part i’, by d. basu.
\newblock {\em Foundations of statistical inference\/}, {\bf 326}.

\bibitem[Kalbfleisch and Prentice(2002)Kalbfleisch and Prentice]{kalbfleisch2002statistical}
Kalbfleisch, J.~D. and Prentice, R.~L. (2002).
\newblock {\em The statistical analysis of failure time data\/}.
\newblock John Wiley \& Sons.

\bibitem[Knorr-held(2000)Knorr-held]{knorr2000bayesian}
Knorr-held, L. (2000).
\newblock Bayesian modeling of inseparable space–time variation in disease risk.
\newblock {\em Statistics in Medicine\/}, {\bf 19}, 2555--2567.

\bibitem[Lewin(2009)Lewin]{lewin2009access}
Lewin, K.~M. (2009).
\newblock Access to education in {sub-Saharan Africa}: patterns, problems and possibilities.
\newblock {\em Comparative Education\/}, {\bf 45}, 151--174.

\bibitem[Lumley(2024)Lumley]{survey}
Lumley, T. (2024).
\newblock survey: analysis of complex survey samples.
\newblock R package version 4.4.

\bibitem[Lutz {\em et~al.}(2008)Lutz, Cuaresma, and Sanderson]{lutz2008demography}
Lutz, W., Cuaresma, J.~C., and Sanderson, W. (2008).
\newblock The demography of educational attainment and economic growth.
\newblock {\em Science\/}, {\bf 319}, 1047--1048.

\bibitem[Macharia {\em et~al.}(2023)Macharia, Moturi, Mumo, Giorgi, Okiro, Snow, and Ray]{macharia2023modelling}
Macharia, P.~M., Moturi, A.~K., Mumo, E., Giorgi, E., Okiro, E.~A., Snow, R.~W., and Ray, N. (2023).
\newblock Modelling geographic access and school catchment areas across public primary schools to support subnational planning in {Kenya}.
\newblock {\em Children's Geographies\/}, {\bf 21}, 832--848.

\bibitem[Manda and Meyer(2005)Manda and Meyer]{manda2005age}
Manda, S. and Meyer, R. (2005).
\newblock Age at first marriage in malawi: a bayesian multilevel analysis using a discrete time-to-event model.
\newblock {\em Journal of the Royal Statistical Society Series A: Statistics in Society\/}, {\bf 168}, 439--455.

\bibitem[Meara {\em et~al.}(2008)Meara, Richards, and Cutler]{meara2008gap}
Meara, E.~R., Richards, S., and Cutler, D.~M. (2008).
\newblock The gap gets bigger: changes in mortality and life expectancy, by education, 1981--2000.
\newblock {\em Health affairs\/}, {\bf 27}, 350--360.

\bibitem[{Ministry of Education, Tanzania}(2017){Ministry of Education, Tanzania}]{TanzaniaESDP2020}
{Ministry of Education, Tanzania} (2017).
\newblock Education sector development plan {(ESDP)} 2016/17 – 2020/21: Delivering quality education and training to all {Tanzanians}.

\bibitem[Riebler {\em et~al.}(2016)Riebler, S{\o}rbye, Simpson, and Rue]{riebler2016an}
Riebler, A., S{\o}rbye, S., Simpson, D., and Rue, H. (2016).
\newblock An intuitive {B}ayesian spatial model for disease mapping that accounts for scaling.
\newblock {\em Statistical Methods in Medical Research\/}, {\bf 25}, 1145--1165.

\bibitem[Rue {\em et~al.}(2009)Rue, Martino, and Chopin]{rue2009approximate}
Rue, H., Martino, S., and Chopin, N. (2009).
\newblock Approximate {B}ayesian inference for latent {G}aussian models using integrated nested {L}aplace approximations (with discussion).
\newblock {\em Journal of the Royal Statistical Society, Series B\/}, {\bf 71}, 319--392.

\bibitem[Simpson {\em et~al.}(2017)Simpson, Rue, Riebler, Martins, and S{\o}rbye]{simpson2017penalising}
Simpson, D., Rue, H., Riebler, A., Martins, T., and S{\o}rbye, S. (2017).
\newblock Penalising model component complexity: A principled, practical approach to constructing priors.
\newblock {\em Statistical Science\/}, {\bf 32}, 1--28.

\bibitem[Singer and Willett(1993)Singer and Willett]{singer1993it}
Singer, J.~D. and Willett, J.~B. (1993).
\newblock It's about time: Using discrete-time survival analysis to study duration and the timing of events.
\newblock {\em Journal of Educational Statistics\/}, {\bf 18}, 155--195.

\bibitem[Smits and Permanyer(2019)Smits and Permanyer]{smits2019subnational}
Smits, J. and Permanyer, I. (2019).
\newblock The subnational human development database.
\newblock {\em Scientific Data\/}, {\bf 6}, 1--15.

\bibitem[Stevens {\em et~al.}(2015)Stevens, Gaughan, Linard, and Tatem]{stevens2015disaggregating}
Stevens, F.~R., Gaughan, A.~E., Linard, C., and Tatem, A.~J. (2015).
\newblock Disaggregating census data for population mapping using random forests with remotely-sensed and ancillary data.
\newblock {\em PLOS ONE\/}, {\bf 10}, 1--22.

\bibitem[Suresh {\em et~al.}(2022)Suresh, Severn, and Ghosh]{suresh2022survival}
Suresh, K., Severn, C., and Ghosh, D. (2022).
\newblock Survival prediction models: an introduction to discrete-time modeling.
\newblock {\em BMC Medical Research Methodology\/}, {\bf 22}, 207.

\bibitem[Thomson {\em et~al.}(2014)Thomson, Shitole, Shitole, Sawant, Subbaraman, Bloom, and Patil-Deshmukh]{thomson2014system}
Thomson, D.~R., Shitole, S., Shitole, T., Sawant, K., Subbaraman, R., Bloom, D.~E., and Patil-Deshmukh, A. (2014).
\newblock A system for household enumeration and re-identification in densely populated slums to facilitate community research, education, and advocacy.
\newblock {\em PLOS ONE\/}, {\bf 9}, 1--9.

\bibitem[{UIS}(2025){UIS}]{UIS}
{UIS} (2025).
\newblock {UIS Stat: SDG 4 Regional Averages}.

\bibitem[UNESCO(2024)UNESCO]{GEM2024}
UNESCO (2024).
\newblock {Global Education Monitoring Report 2024/5}.
\newblock Technical report, UNESCO.

\bibitem[{UNESCO Institute for Statistics} and {African Union}(2024){UNESCO Institute for Statistics} and {African Union}]{UNESCO2024}
{UNESCO Institute for Statistics} and {African Union} (2024).
\newblock 2024 spotlight on basic education completion and foundational learning in {Africa}.
\newblock Accessed: 2025-01-27.

\bibitem[{United Nations}(2015){United Nations}]{SDG4}
{United Nations} (2015).
\newblock {Transforming Our World: The 2030 Agenda for Sustainable Development}.

\bibitem[{United Nations Children's Fund (UNICEF)}(2019){United Nations Children's Fund (UNICEF)}]{unicef2019education}
{United Nations Children's Fund (UNICEF)} (2019).
\newblock {\em Every Child Learns: UNICEF Education Strategy 2019–2030\/}.
\newblock UNICEF, New York.

\bibitem[Wakefield {\em et~al.}(2019)Wakefield, Fuglstad, Riebler, Godwin, Wilson, and Clark]{wakefield2019estimating}
Wakefield, J., Fuglstad, G.-A., Riebler, A., Godwin, J., Wilson, K., and Clark, S.~J. (2019).
\newblock Estimating under-five mortality in space and time in a developing world context.
\newblock {\em Statistical Methods in Medical Research\/}, {\bf 28}, 2614--2634.

\bibitem[Willett and Singer(1991)Willett and Singer]{willett1991whether}
Willett, J.~B. and Singer, J.~D. (1991).
\newblock From whether to when: New methods for studying student dropout and teacher attrition.
\newblock {\em Review of Educational Research\/}, {\bf 61}, 407--450.

\bibitem[Wu and Wakefield(2024)Wu and Wakefield]{wu2024modelling}
Wu, Y. and Wakefield, J. (2024).
\newblock Modelling urban/rural fractions in low- and middle-income countries.
\newblock {\em Journal of the Royal Statistical Society Series A: Statistics in Society\/}, {\bf 187}, 811--830.

\end{thebibliography}

\newpage 
\setcounter{section}{0}
\setcounter{figure}{0}
\setcounter{equation}{0}

\makeatletter
\renewcommand \thesection{S\@arabic\c@section}
\renewcommand \thefigure{S\@arabic\c@figure}
\renewcommand \theequation{S\@arabic\c@equation}

\makeatother

\centerline{{\huge Supplemental Materials} }

\section{Derivation of Standard Error for Years of Education}
\label{sec:supp-derive-delta-method}
In this section, we derive the delta method for estimating the survival probabilities and years of education under the survey GLM framework. The model parameters, $\bbeta$, represent the logit of hazards and are transformed into the quantities of interest based on Equations (\ref{eq:educ-year-from-surv}) and (\ref{eq:educ-year-surv}) in the method section. The derivatives required for the delta method can be obtained through the chain rule and are analytically tractable. Delta method in the modified weighted estimation method can be derived directly from the hazard, following similar and simpler steps; thus, we focus here on the survey GLM case.

To simplify the derivation, we adopt the strategy of repetitively fitting the survey GLM by treating each domain as the reference domain. In this reparametrization, the survival function (school completion probabilities) and expected years of education depend solely on the model parameters $\beta$s (logit hazards) and not on the domain-specific effects $\gamma$s, making the delta method derivation more straightforward. Such reparametrizations, although differing in their choice of reference domain, are statistically equivalent because the fitted domain-specific survival functions are invariant to the parametrization. 

\subsection*{Delta Method for Survival Function}

In the survey GLM model, the parameters $\boldsymbol{\beta} = \{\beta_k\}$ represent the logit-transformed hazards for each grades (for the reference domain), where 
\[
\beta_k = \text{logit}(h_k) = \log\left(\frac{h_k}{1 - h_k}\right).
\]
The survival function \( S(t) \) is then expressed in terms of the parameters $\bbeta$ as:
\[
S(t) = \prod_{k=0}^{t-1} \left( 1 - \text{expit}(\beta_k) \right), \quad t = 1,2,\ldots,K,
\]

The maximum likelihood estimator (MLE) of $\boldsymbol{\beta}$, denoted by $\widehat{\boldsymbol{\beta}}$, is asymptotically normal:
$
\widehat{\boldsymbol{\beta}} \sim N\left(\boldsymbol{\beta}, \widehat{\Sigma}\right),
$
where $\widehat{\Sigma}$ is the estimated covariance matrix of $\widehat{\boldsymbol{\beta}}$, obtained from the \texttt{svyglm()} function in the \texttt{survey} package.

Using the delta method, the asymptotic variance of the estimated survival probability, \(\widehat{S(t)}\), is given by:

\begin{equation}
\label{eq:delta-surv-prob}
\widehat{\text{Var}}\left(\widehat{S(t)}\right) = \left( \frac{\partial S(t)}{\partial \boldsymbol{\beta}} \bigg|_{\boldsymbol{\beta} = \widehat{\boldsymbol{\beta}}} \right)^\top 
\cdot \widehat{\Sigma} \cdot 
\left( \frac{\partial S(t)}{\partial \boldsymbol{\beta}} \bigg|_{\boldsymbol{\beta} = \widehat{\boldsymbol{\beta}}} \right),
\end{equation}

The gradient vector of \(S(t)\) with respect to \(\boldsymbol{\beta}\) is:
\[
\frac{\partial S(t)}{\partial \beta_j} =
\begin{cases} 
0, & j \geq t, \\
-S(t) \cdot \text{expit}(\beta_j), & j < t.
\end{cases}
\]

For \(j < t\), the gradient is derived as:
\begin{align*}
\frac{\partial S(t)}{\partial \beta_j} 
&= \frac{S(t)}{1 - \text{expit}(\beta_j)} \cdot \frac{\partial}{\partial \beta_j}(1 - \text{expit}(\beta_j)) \\
&= \frac{S(t)}{1 - \text{expit}(\beta_j)} \cdot \left(-\text{expit}(\beta_j) \cdot (1 - \text{expit}(\beta_j)) \right) \\
&= -S(t) \cdot \text{expit}(\beta_j).
\end{align*}

Equation (\ref{eq:delta-surv-prob}) enables the calculation of confidence intervals for \(S(t)\).

\subsection*{Delta Method for Expected Years of Education}

The expected years of education, denoted by $\mu$ is computed as the sum of survival probabilities:
\begin{equation}
\mu = \sum_{t=1}^{K} S(t).
\end{equation}

Extending the delta method for the survival function, the asymptotic variance of the estimated years of education, \(\hat{\mu}\), is given by:

\begin{equation}
\label{eq:delta-educ-years}
\widehat{\text{Var}}\left(\hat{\mu}\right) = \left( \frac{\partial \mu}{\partial \boldsymbol{\beta}} \bigg|_{\boldsymbol{\beta} = \widehat{\boldsymbol{\beta}}} \right)^\top 
\cdot \widehat{\Sigma} \cdot 
\left( \frac{\partial \mu}{\partial \boldsymbol{\beta}} \bigg|_{\boldsymbol{\beta} = \widehat{\boldsymbol{\beta}}} \right),
\end{equation}

where $\widehat{\Sigma}$ is the estimated covariance matrix of $\widehat{\boldsymbol{\beta}}$, and the gradient of \( \mu \) with respect to $\bbeta$ is:
\begin{equation*}
\frac{\partial \mu}{\partial \beta_j} = 
\sum_{t=j+1}^{K} \frac{\partial S(t)}{\partial \beta_j}.
\end{equation*}

Substituting the gradient of \( S(t) \) yields:
\begin{equation*}
\frac{\partial \mu}{\partial \beta_j} = 
-\sum_{t=j+1}^{K} S(t) \cdot \text{expit}(\beta_j).
\end{equation*}

\subsection*{Confidence Intervals}
Based on the Equation (\ref{eq:delta-surv-prob}) and (\ref{eq:delta-educ-years}), the confidence intervals for school completion probabilities and years of education can be calculated as:
\begin{equation*}
\widehat{S(t)} \pm z_{\alpha/2} \cdot \sqrt{\widehat{\text{Var}}(\widehat{S(t)})},
\end{equation*}
\begin{equation*}
\hat{\mu} \pm z_{\alpha/2} \cdot \sqrt{\widehat{\text{Var}}(\hat{\mu})}.
\end{equation*}

\newpage 

\section{Urban/Rural Stratification} 
\label{sec:supp-UR-stratification}

In spatial models, we need to adjust for urban and rural when the outcome depends on U/R and the survey carried out unequal sampling with respect to urban and rural clusters. The pipeline for properly accounting for this stratification is summarized in Figure \ref{fig:UR-pipeline-overview}. As discussed in the main manuscript, obtaining U/R specific estimates is straightforward by fitting discrete-time survival models separately for urban and rural areas. The overall estimate for UYS or the education trajectory $S(t)$ is then computed as a weighted combination of these estimates, with the urban fraction $r_i$ as the weight. In a Bayesian framework, this aggregation naturally accounts for uncertainty, as posterior samples for the overall estimate are derived by combining posterior samples from the U/R specific models.

\tikzstyle{decision} = [diamond, draw, fill=blue!20, 
    text width=5.5em, text badly centered, node distance=3cm, inner sep=0pt]
\tikzstyle{block} = [rectangle, draw, fill=blue!20, 
    text width=10em, text centered, rounded corners, minimum height=4em]
\tikzstyle{line} = [draw, -latex']
\tikzstyle{cloud} = [draw, ellipse,fill=red!20, node distance=3cm, text width=4.5em,text centered,
    minimum height=2em]
    
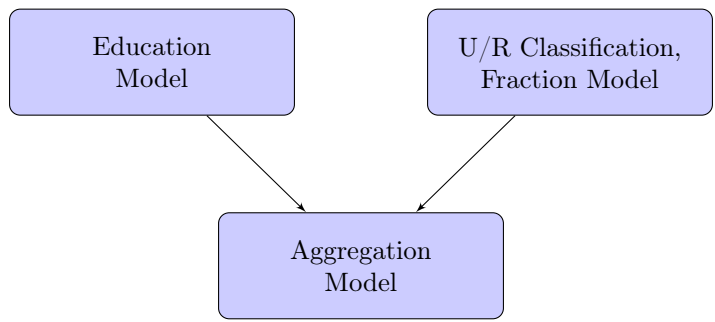
\begin{figure} [!ht]

     \begin{center}
    \begin{small}
    
\begin{tikzpicture}[node distance = -.3cm, auto]
    \node [block] (level1left) {Education\\ Model};
        \node [block, xshift=5.5cm] (level1right) {U/R Classification, \\ Fraction Model};
    \node [block, below of=level1left, yshift=-3cm, xshift=2.75cm] (level2) {Aggregation\\ Model};

    \path [line] (level1left) -- (level2);
        \path [line] (level1right) -- (level2);
   
\end{tikzpicture}

    \caption{Overview of modeling process}

\label{fig:UR-pipeline-overview}

\end{small}
     \end{center}
\end{figure}

A key challenge in this approach is accurately estimating the urban fraction $r_i$, which represents the proportion of the target population classified as urban in the survey’s sampling frame. Unsurprisingly, direct estimation at fine geographic scales (e.g., Admin-2) is unfeasible due to data sparsity. To address this, we adopt the pipeline introduced in \citet{wu2024modelling}, with necessary simplifications.

The core idea is to represent sampling clusters as pixels on a gridded surface for the target area. We develop a classification model to label each pixel as urban or rural and use the WorldPop population density \citep{stevens2015disaggregating} to allocate the relevant population to these pixels. This enables the estimation of the urban fraction within any given area.

Formally, let $V$ denote an urban classification map where each pixel is labeled as urban (1) or rural (0), and a generic population density map $H$ for the study region. For a pixel indexed by $g$ with coordinates $\bms(g)$, we denote its urban/rural status and population density as $V_{\scaleto{\bms}{4pt}(g)}\equiv v_g$ and $H_{\scaleto{\bms}{4pt}(g)}\equiv H_g$ respectively. Then urban fraction in a generic area can then be computed as, 
\begin{equation}
\label{eq:supp-ur-frac}
    r =  \frac{\sum v_g \times  H_{g}}{\sum H_{g}}
\end{equation}

\subsection*{Application to 2022 Tanzania DHS}
To estimate the urban fraction for females in each age group, as required for the analysis based on 2022 Tanzania DHS, we follow these steps:
\begin{enumerate}
\item \textbf{Identify the training data.} Since we aim to construct a U/R surface consistent with the sampling frame, we include clusters from multiple DHS surveys that share the same sampling frame, which contain the reported geographical coordinates and U/R status of each cluster. For our purpose, both 2022 Tanzania DHS and 2015--2016 Tanzania DHS use the 2012 Tanzania Population and Housing Census as their sampling frame. We therefore combine clusters from these two surveys and define them as the training set.

\item  \textbf{Build the U/R classification model.} We incorporate multiple covariates into our classification model to predict urban/rural (U/R) status. We obtain the total population raster for Tanzania (2012) from WorldPop \citep{stevens2015disaggregating}, and we include nighttime lights data from National Oceanic and Atmospheric Administration (NOAA), a highly predicative indicator of urbanization due to its strong correlation with human settlements and infrastructure. To account for regional differences in urban classification patterns, we also include Admin-1 region as a categorical predictor. For model training, we extract covariate values at the geographical coordinates of DHS survey clusters, whose known U/R label serves as the response variable. Using these extracted features, we train a Gradient Boosted Trees (GBT) model to classify pixels into urban and rural.

\item \textbf{Transform predicted probabilities into a binary U/R label surface.} The classification model outputs probabilities of urban classification for each pixel. To ensure consistency with the survey-defined U/R classification, we apply a thresholding algorithm: pixels with predicted probabilities above a chosen threshold are classified as urban (1), while the rest are labeled rural (0). The threshold is calibrated against known urban proportions for the total population at the Admin-1 level to align with the sampling frame for the survey.

\item \textbf{Compute urban fractions for target population/area.} To compute urban fractions for our analysis for 2022 Tanzania DHS survey, we obtain population density maps for females in 5-year age groups from WorldPop. Using the calibrated urban classification map and the pixel-level distribution of the relevant sub-population in the target region $i$, we compute urban fractions $r_i$ based on Equation (\ref{eq:supp-ur-frac}).

\end{enumerate}

\newpage 

\section{Additional Details on Simulation Study} 
\label{sec:supp-sim-setup}

\subsection{Education Trajectory and Right-Censoring}
\label{sec:supp-educ-history-visual}

Recall Equation (\ref{eq:cal-exit-time}) from the main manuscript. For each individual, school exit time (completion or discontinuation) is determined as:
\[
T_{\text{exit}} = T_{\text{birth}} + E + T,
\]
where \( T_{\text{birth}} \) is the birth year, \( E \) is the school entrance age, and \( T \) is UYS. Right-censoring was determined by comparing \( T_{\text{exit}} \) with the survey time \( T_{\text{survey}} \).

\definecolor{lightblue}{RGB}{173,216,230}  

\newcommand{\hollowcircle}{\tikz[baseline=-0.4ex] \draw (0,0) circle (0.6ex);}  
\newcommand{\bluesolid}{\tikz[baseline=-0.5ex] \draw[thick, blue] (0,0) -- (1.5em,0);} 
\newcommand{\lightbluedashed}{\tikz[baseline=-0.5ex] \draw[thick, lightblue, dashed] (0,0) -- (1.5em,0);} 
\newcommand{\greendashed}{\tikz[baseline=-0.5ex] \draw[thick, green, dashed] (0,0) -- (1.5em,0);} 
\newcommand{\reddashed}{\tikz[baseline=-0.5ex] \draw[thick, red, dashed] (0,0) -- (1.5em,0);} 
\newcommand{\filledsquare}{\tikz[baseline=-0ex] \fill[black] (0,0) rectangle (1.2ex,1.2ex);} 
\newcommand{\hollowsquare}{\tikz[baseline=0ex] \draw[] (0,0) rectangle (1.2ex,1.2ex);} 
\newcommand{\triangleup}{\tikz[baseline=-0.1ex] \draw[thick,gray] (0,0) -- (0.5ex,1ex) -- (1ex,0) -- cycle;} 

\begin{figure} [!ht]
    \centering
    \includegraphics[clip, trim=0cm 0cm 0cm 0cm, width=0.99\linewidth]{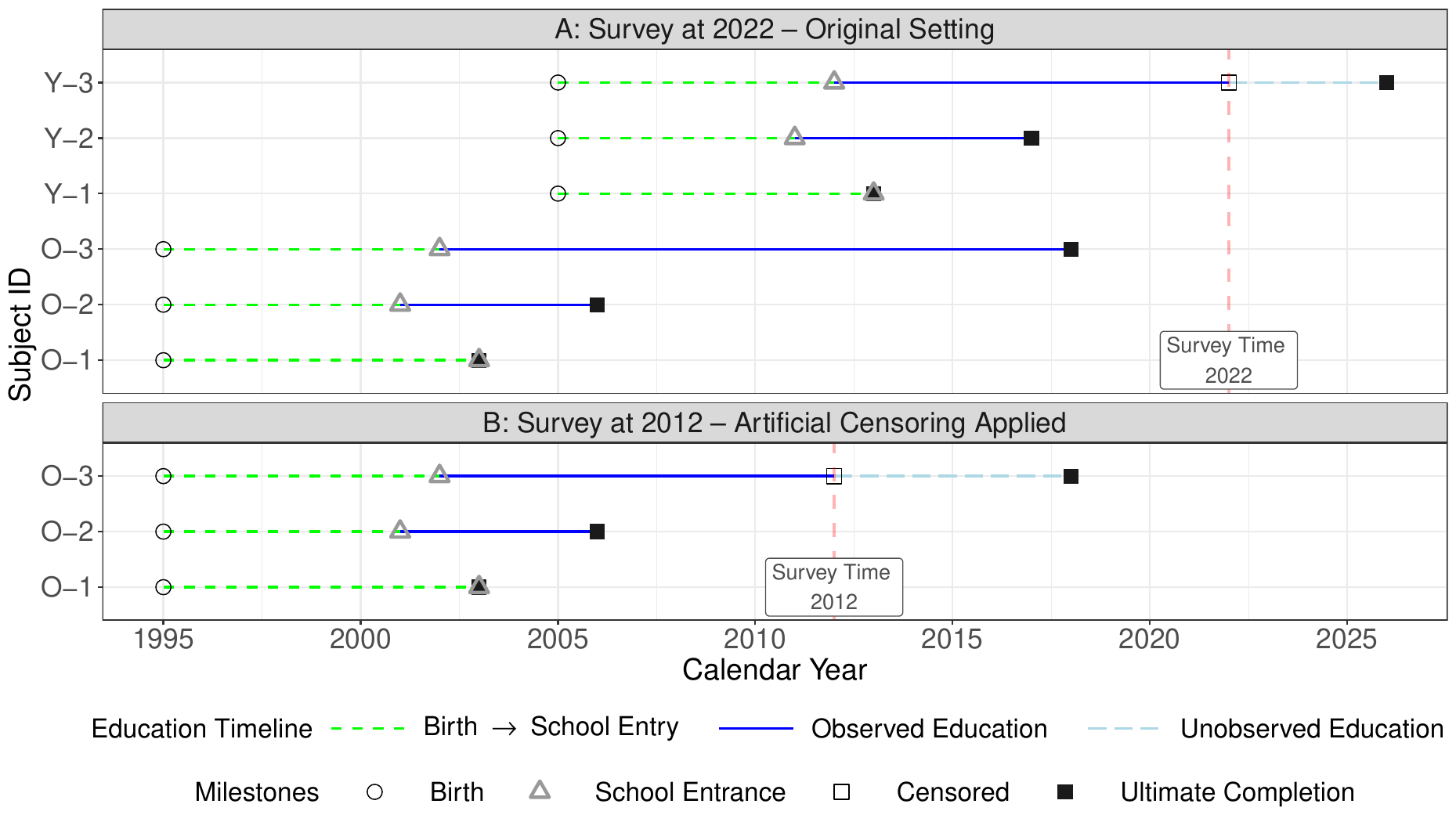}
    \caption{Illustration of the framework for modeling and decomposing individuals' education trajectories.} 
    \label{fig:supp-educ-history-visual}
\end{figure}

Figure \ref{fig:supp-educ-history-visual} illustrates how education history is decomposed and modeled. Key components include:
\begin{itemize}
    \item \hollowcircle{} denotes the birth year (\(T_{\text{birth}}\)).
    \item \greendashed{} represents period from birth to school entry (\(E\)), with school entrance time denoted by \triangleup{}.
    \item \bluesolid{} (observed education) and \lightbluedashed (unobserved education) constitutes of UYS (\(T\)), with ultimate completion time denoted by \filledsquare{}.
    \item \reddashed{} indicates survey time (\(T_{\text{survey}}\)), with censorship denoted by \hollowsquare{}.
\end{itemize}

The figure presents two scenarios:

\begin{enumerate}
    \item Scenario A: Survey in 2022 (Original Setting)  
    This scenario reflects real-world data collection. Younger cohorts (Y-1, Y-2, Y-3, ages 15--19 in 2022) are closer to the survey date, making them more susceptible to censoring. For example, Y-3 is still in school at the time of the survey, her education trajectory is right-censored. In contrast, Y-1 and Y-2 have completed their education before the survey, meaning their full education history is observed. Y-1 has no education, so her school entrance age effectively overlaps with her ultimate completion. Older cohorts (O-1, O-2, O-3, ages 25--29 in 2022) have fully completed their education before the survey. Even those with high educational attainment (e.g., O-3, who attained tertiary education) are not subject to censoring.

    \item Scenario B: Survey in 2012 (Artificial Censoring Applied). This scenario simulates an earlier survey, shifting data collection back by 10 years. Right-censoring is introduced to mimic the conditions younger cohorts would face in a real-world setting. Individuals like O-3---who would have continued to receive education beyond 2012---are now censored, as their full education trajectory would not have been captured if the survey had occurred earlier.
\end{enumerate}

\subsection{Modeling Delayed Entrance}

Let \( T \) denote the years of education completed by an individual, and let $A_s$ be their age at the time of the survey. We focus on individuals who have received at least some formal education, i.e., \( T > 0 \). The education history of an individual can be traced back to the age \( E \) at which they started the first grade.

Similar to the education years model, we assume no grade repetition or gaps in education. For individuals still in school at the time of the survey (\( \delta = 0 \)), the exact entrance age can be determined as \( E = A_s - T \). For individuals who have discontinued school (\( \delta = 1 \)), the exact entrance age is unknown, but partial information is available: \( E \leq A_s - T \), as a later entrance age would imply they were still in school at the time of the survey.

Thus, the observed entrance age \( E \) is left-censored, in contrast to the right-censoring in years of education. Interestingly, exactly one of \( E \) and \( T \) is censored for any individual. To reflect this, we define the censoring indicator for \( E \) as \( \delta^* = 1 - \delta \).

We treat the entrance age \( E \) in a discrete-time format, similar to the formulation for UYS. While birth dates are recorded in century month code (CMC), allowing \( E \) to be derived at a monthly resolution, we model only integer years, denoted as \(\lfloor E \rfloor \). This is because data sparsity does not support monthly-level modeling, and since the goal is to model school start age distribution for simulation, sampling at a monthly scale may misalign birth month with school start month (for example, January in Tanzania). The method for refining \( E \) to a finer resolution is detailed in the next section.

To further justify the model setup, we assume that entrance ages fall within the range of 5--15 years, since starting school outside this range is exceptionally rare, as confirmed by our analysis of the 2022 Tanzania DHS.

To facilitate modeling, we introduce a transformed variable \( E^* = 15 - \lfloor E \rfloor  \), which maintains a one-to-one relationship with \( \lfloor E \rfloor  \). Estimating the distribution of \( E^* \) directly translates to an estimate for \( \lfloor E \rfloor  \). By design, \( E^* \) is right-censored, making it naturally suited for discrete-time survival modeling. Its censoring status aligns with that of \( T \). 

The transformed entrance age \( E^* \) can be interpreted as the \textit{reverse entrance age}. Under the discrete-time survival model setting, an individual with an event at \( E^* = 9 \),for example, implies they did not start school from ages [15--14),...,[7--6), and started at age 6 (\( \lfloor E \rfloor  = 6 \iff E^* = 9 \)).

The methodology developed for years of education applies directly to this reverse-engineered entrance age. We can use the modified weighted estimation method to estimate the distribution of \( E^* \). Note that for individuals aged 30 or older at the time of the survey, \( \lfloor E \rfloor  \) is always censored, as it is highly unlikely for such individuals to still be in school. Consequently, entrance age estimates based on a DHS survey are only meaningful for those aged 15--29 at the time of the survey.

We can transform the estimated distribution of  \( E^* \) back to 
$\lfloor E \rfloor $ and simulate the school entrance age accordingly. While the actual distribution of school entrance ages may have varied historically, this serves as a reasonable approximation for our purposes. Importantly, the validity of our method for addressing censoring in years of education is robust to the specific form of the entrance age distribution, given that censoring mechanism is non-informative (as in the case of an end-of-study censorship occurring here).

\subsection{Additional Details on Simulation Setup}

Recall Equation (\ref{eq:cal-exit-time}) from the main manuscript, for each individual, school exit time (discontinuation/completion) can be determined as:
\[
T_{\text{exit}} = T_{\text{birth}} + E + T,
\]
where \( T_{\text{birth}} \) is the birth year, \( E \) is the school entrance age, and \( T \) is the observed years of completed education.

A key consideration arises regarding the time resolution in the available data. Birth dates are recorded in Century Month Code (CMC), and the academic year in Tanzania begins in January, also recorded in CMC. However, the school entrance age sampled from the delayed entrance model is available only as integer years, and we denote this as \(\lfloor E \rfloor \). 

To introduce a finer resolution, we decompose the entrance age as:
\[
E=\lfloor E \rfloor + \{E\}
\]
\noindent where \(\lfloor E \rfloor \) represents the entrance age in completed years, and \(\{E\}\) is the additional fraction of a year that accounts for the exact school start time.

To determine \(\{E\}\), we align school entry with the first school start time after the individual turns \(\lfloor E \rfloor \) years old. For example, if an individual is born in April 2000 and is recorded to start school at age \(\lfloor E \rfloor  = 7\), their actual school entry occurs in January 2008---the first school start time following their 7th birthday (April 2007). In this case, the fractional term is: $\{E\} = \frac{9}{12}$, meaning the individual is 7 years and 9 months old at school entry. This adjustment ensures that school entrance timing is properly aligned with the education system’s structure while maintaining consistency with the available data resolution.

Another subtlety in the simulation study is that dropout cannot be sampled within a grade interval because DHS surveys do not record incomplete years of education such that we have no data to model this within-grade dropout pattern. This does not affect our estimate for UYS since our target measure is completed years, but it introduces a small difference in conducting the simulation.

In real-data analysis, dropout can occur within a school year. If an individual is still in school at the time of the survey, we handle censoring by excluding their partially completed final year, as their eventual completion status is unknown.

In contrast, in the simulation, education data is available only in completed years, meaning that dropout can only occur at the end of a grade level. If an individual is censored in the simulation, their most recent school year should be assumed to be completed, since the data does not allow for mid-year dropout. To properly account for this, we postpone the censoring point to the end of the current school year, and record the current grade as completed. 

Although the simulation cannot fully replicate real-world dropout patterns, it effectively captures the overall impact of censoring on completed years of education.

\newpage

\section{Derivation of Beta-Binomial}
\label{sec:supp-beta-bin-derive}

Following Equation (\ref{eq:bernoulli-parameterization}), the education history of an individual
can be formulated into a series of Bernoulli outcomes based on grade intervals, 

\begin{equation*}
\log L_j =\sum_{k=0}^{t_j-(1-\delta_j)} \left[
z_{jk} \log(h_{jk}) + (1 - z_{jk}) \log(1 - h_{jk})
\right],
\end{equation*}

We first focus on a single grade interval $k \in c(0,t_j-(1-\delta_j))$ such that we have a Bernoulli outcome $Z_{jk}\sim\text{Bern}(h_{jk})$.

To incorporate within-cluster correlation in the dropout hazard, we introduce overdispersion by treating the hazard itself as random. Specifically, we replace $h_{jk}$ with $q_{jk}$, denoting the individual-specific random hazard:
\begin{align*}
q_{jk} & \sim \text{Beta}(\alpha_k, \beta_k) \\
Z_{jk}|q_{jk} &\sim \text{Bernoulli}(q_{jk})
\end{align*}
While now $h_{jk}=E(q_{jk})$ refers to as the mean hazard to align with the structure of other models.

The marginal likelihood for $Z_{jk}$ then can be obtained by integrating out $q_{jk}$ as:
\begin{align*}
P(Z_{jk} = z_{jk}) &= \int_0^1 q_{jk}^{z_{jk}} (1 - q_{jk})^{(1 - z_{jk})} \frac{q_{jk}^{\alpha_k - 1} (1 - q_{jk})^{\beta_k - 1}}{B(\alpha_k, \beta_k)} \, dq_{jk} \nonumber \\
&= \frac{1}{B(\alpha_k, \beta_k)} \int_0^1 q_{jk}^{z_{jk} + \alpha_k - 1} (1 - q_{jk})^{(1 - z_{jk}) + \beta_k - 1} \, dq_{jk}  \nonumber \\
& = \frac{B(z_{jk} + \alpha_k, (1 - z_{jk}) + \beta_k)}{B(\alpha_k, \beta_k)},
\end{align*}

\noindent which we recognize as a \textbf{beta-Bernoulli distribution} with mean $\frac{\alpha_k}{\alpha_k+\beta_k}=E(q_{jk})=h_{jk}$.

Now, for a group of individual that shares the same characteristic (same mean hazard $h_k$) within a cluster, we sum over their event history for grade interval $k$. Let $n_k$ be the total number of individuals in the risk set at grade interval $k$, i.e.,~passed through previous grade intervals. Let $Y_k$ denote the number of individuals who discontinue education in this interval, such that: 
$Y_k = \sum_{j=1}^{n_k} Z_{jk}$.

Since each individual’s dropout follows a beta-Bernoulli distribution with the same hyperparameters \( \alpha_k \) and \( \beta_k \) and assumed to be independent, the total number of dropouts at grade interval \( k \), denoted as \( Y_k \), follows a \textbf{beta-binomial distribution}:
\begin{equation*}
P(Y_k = y_k) = \binom{n_k}{y_k} \frac{B(y_k + \alpha_k, n_k - y_k + \beta_k)}{B(\alpha_k, \beta_k)}.
\end{equation*}

This formulation accounts for within-cluster correlation and the expected number of dropouts and its variance are given by:
\begin{align*}
\mathbb{E}[Y_k] &= n_k \frac{\alpha_k}{\alpha_k + \beta_k} = n_k h_k, \\
\text{Var}(Y_k) &= n_k h_k (1 - h_k) \phi,
\end{align*}

where the overdispersion parameter is
\begin{equation*}
\phi = \frac{1}{1 + (n_k - 1) \rho}, \quad \text{with} \quad \rho = \frac{1}{\alpha_k + \beta_k + 1}.
\end{equation*}

Thus, the dropout count can be written as:
\begin{equation*}
Y_k \mid h_k \sim \text{BetaBinomial}(n_k, h_k, \phi).
\end{equation*}

This formulation enables the introduction of a regression structure by modeling the logit transformation of \( h_k \), linking dropout probabilities to model components as described in Equation (\ref{eq:educ-betabinomial}) and (\ref{eq:educ-cluster-model-mean-admin2}) in the main text.

\newpage 

\section{Proportional Odds Assumption}
\label{sec:supp-prop-odds-assumption}

In this section, we assess the validity of the proportional odds assumption in modeling UYS using a discrete-time survival model. The proportional odds assumption posits that the effects of groups (such as birth cohort, region, or their interactions) remain consistent across grade levels. This implies that the log-odds of conditional dropout probability across grades should exhibit parallel trends across different groups. 

\begin{figure} [!ht]
    \centering
    \includegraphics[clip, trim=0cm 0cm 0cm 0cm, width=1\linewidth]{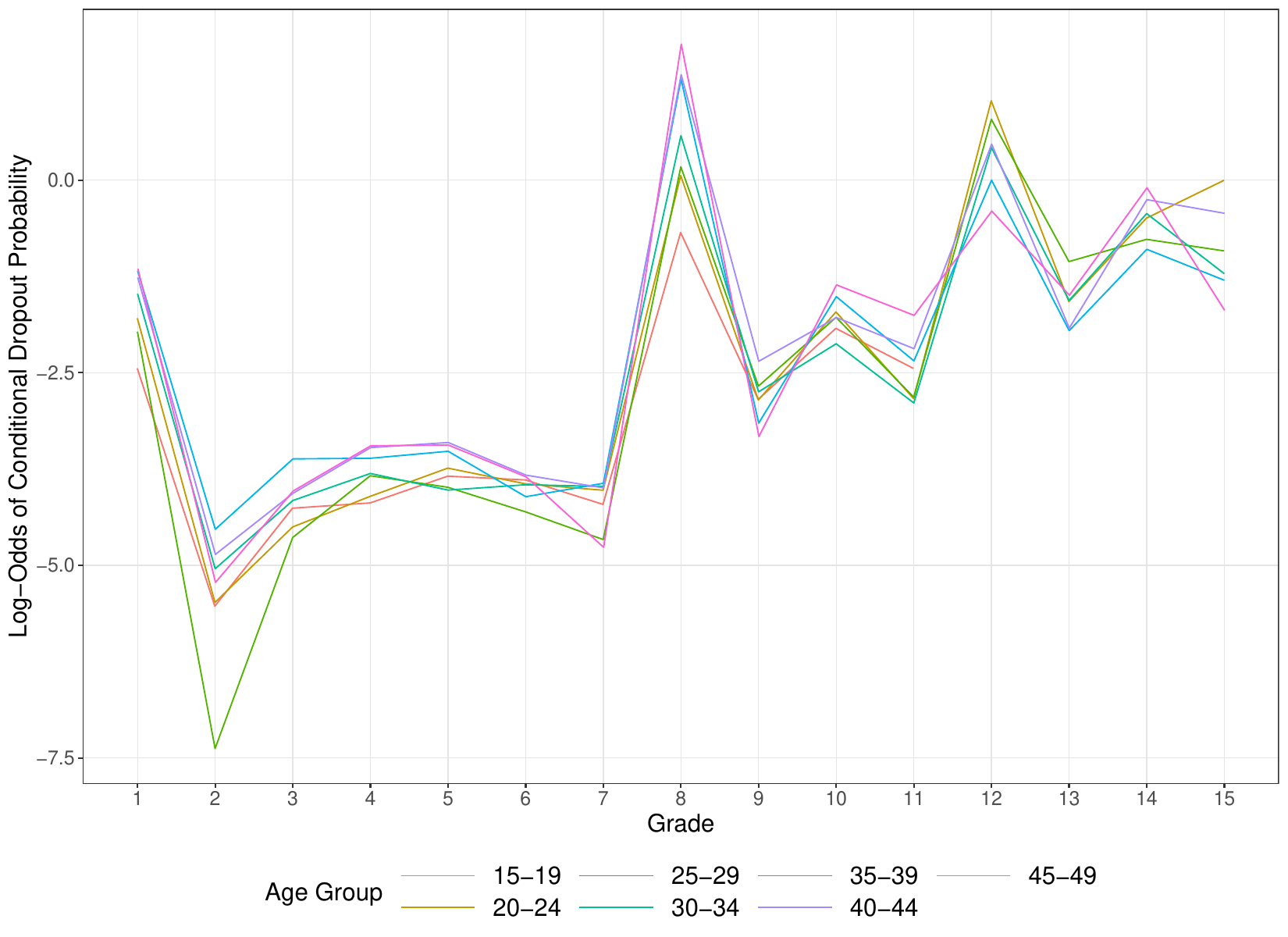}
    \caption{Log-odds of grade-specific conditional dropout probability, stratified by age group.}
    \label{fig:natl-grade-prop-check}
\end{figure}

We begin by examining national-level estimates. Figure \ref{fig:natl-grade-prop-check} presents the log-odds of grade-specific dropout probability for females in the 2022 Tanzania DHS, stratified by age group. For example, the estimate at grade 3 represents the log-odds of the dropout probability within the grade interval [2,3). These estimates are derived through modified weighted estimation. 

Overall, the parallel trends across age groups suggest that the proportional odds assumption holds, meaning that birth cohort effects remain consistent across grade levels.  A potential deviation occurs around grade 12 (upper secondary), where younger cohorts (20--24 and 25--29) show higher dropout probabilities than older cohorts, diverging from the overall pattern. In addition, for the most recent birth cohort (ages 15--19 in 2022), we do not observe data beyond lower secondary education due to end-of-study censorship. 

\begin{figure} [!ht]
    \centering
    \includegraphics[clip, trim=0.5cm 0cm 0.5cm 1cm, width=1\linewidth]{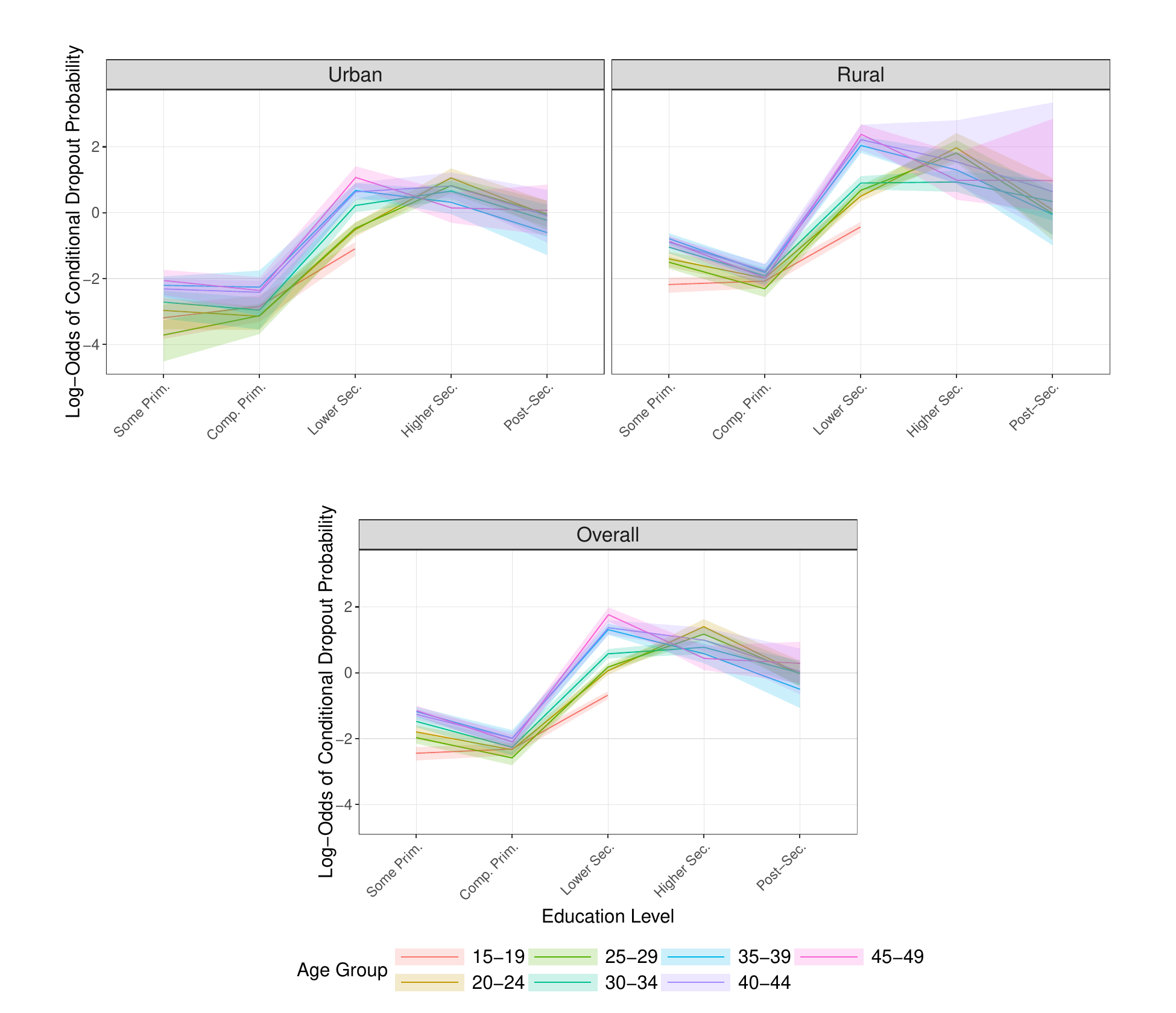}
    \caption{Log-odds of conditional dropout probability at national level across education levels, stratified by age group. The top panels present estimates separately for urban and rural areas, while the bottom panel displays overall estimates.}
    \label{fig:natl-educ-level-prop-check}
\end{figure}

Building on this, Figure \ref{fig:natl-educ-level-prop-check} aggregates dropout probabilities across education levels to further evaluate proportionality. These estimates, also obtained through modified weighted estimation, are plotted along an x-axis representing educational levels:

\begin{itemize}
    \item \textbf{Some Prim.} - Some primary education (1-6 years of education)
    \item \textbf{Comp. Prim.} – Completed primary education  (7 years of education)
    \item \textbf{Lower Sec.} – Lower secondary education (8-11 years of education)
    \item \textbf{Higher Sec.} – Higher secondary education (12-13 years of education)
    \item \textbf{Post-Sec.} – Postsecondary education (14+ years of education)
\end{itemize}

For example, the second category represents the log-odds of failing to complete primary education, conditional on having attended at least grade 1. The results indicate no substantial deviation from the proportional odds assumption, as dropout trends remain largely parallel across age groups. Similarly, urban-rural stratified analyses exhibit similar patterns. Minor deviations are observed --- the 20--24 and 25--29 age groups display slightly elevated dropout risks at the higher secondary level.

\begin{figure} [!ht]
    \centering
    \includegraphics[clip, trim=0cm 0cm 0cm 0cm, width=1\linewidth]{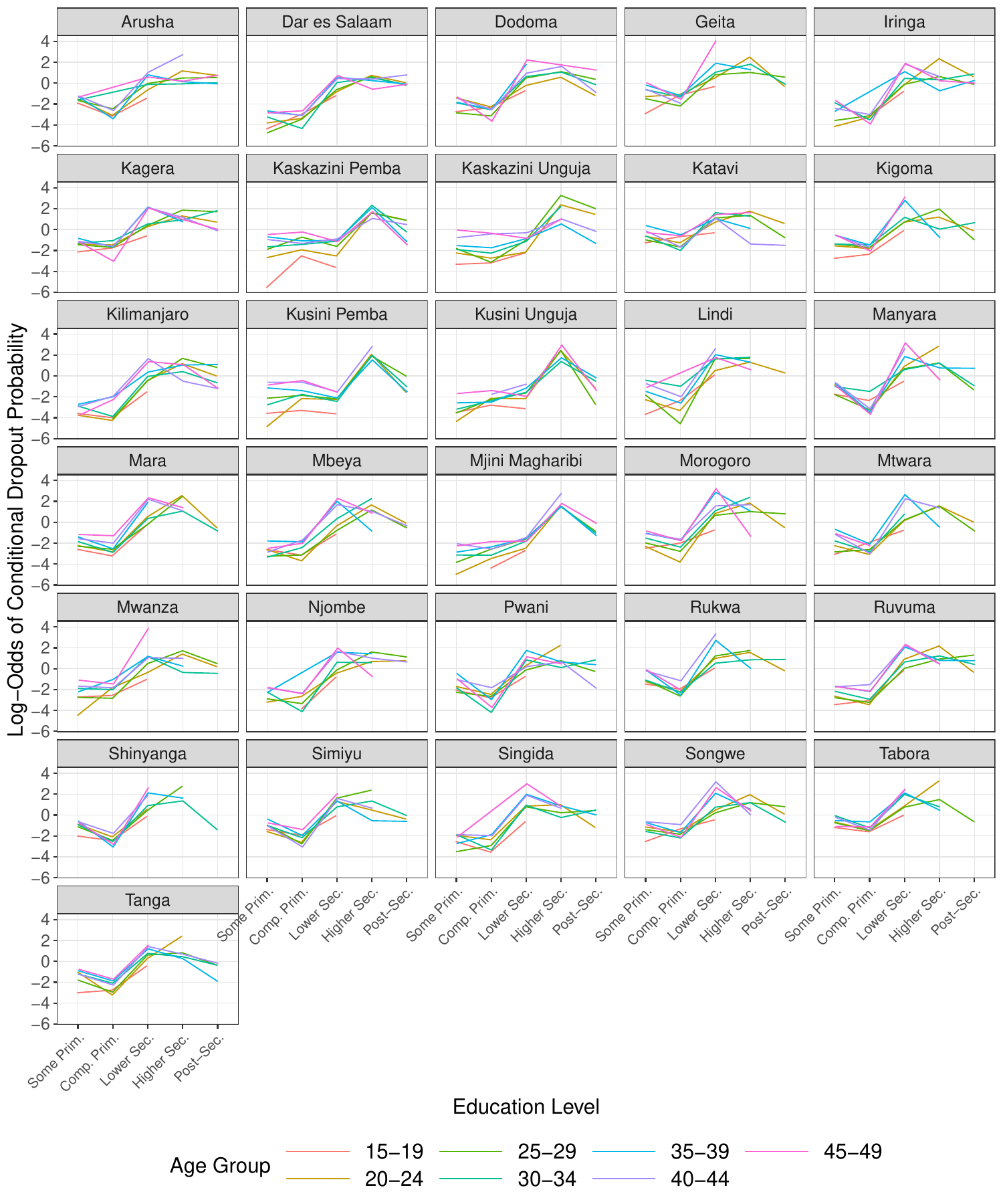}
    \caption{Log-odds of conditional dropout probability across education levels at the Admin-1 level, stratified by age group. }
    \label{fig:adm1-educ-level-prop-check}
\end{figure}

To further examine regional variations, we evaluate the proportional odds assumption at the Admin-1 level. Figure \ref{fig:adm1-educ-level-prop-check} presents log-odds of conditional dropout probability across education levels for different administrative regions. Certain regions display localized deviations, but in general, the trends at the Admin-1 level align with national-level findings, supporting the proportional odds assumption.

Overall, while minor deviations exist at both national and subnational levels, they do not systematically challenge the proportional odds assumption. 

\clearpage
\newpage

\section{Additional Result Visualizations}

\subsection{Urban/Rural Specific National Estimates}
\label{sec:supp-figure-natl-UR}

\begin{figure} [!ht]
    \centering
    \includegraphics[clip, trim=0cm 0cm 0cm 0cm, width=0.99\linewidth]{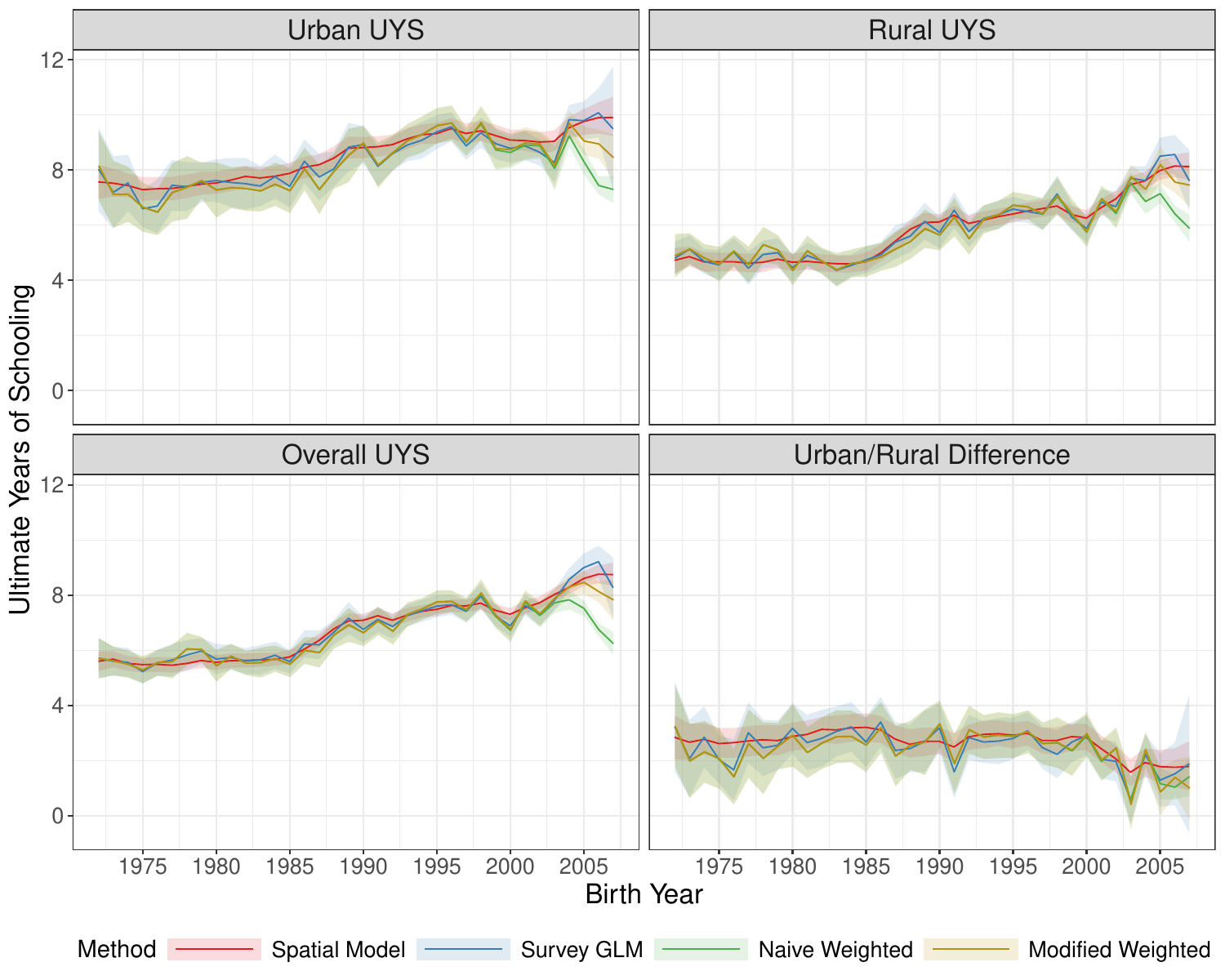}
    \caption{Estimated UYS by birth year for females in the 2022 Tanzania DHS. Urban/rural stratified and overall estimates are presented alongside with urban-rural difference.}
    \label{fig:natl-all-UYS-compare}
\end{figure}

Figure \ref{fig:natl-all-UYS-compare} presents estimates of UYS by birth year for females in the 2022 Tanzania DHS, via different estimation approaches: \textbf{naive weighted estimation}, \textbf{modified weighted estimation}, \textbf{survey GLM}, and \textbf{spatial model}. The four panels present overall trends, separate urban and rural estimates, and the urban-rural difference.

A persistent urban-rural gap is evident, with urban areas consistently achieving higher UYS than rural areas across all birth cohorts. While both urban and rural estimates show a steady upward trend, rural areas exhibit slightly greater improvement over time. Consequently, the urban-rural gap remains stable for older cohorts but narrows slightly in more recent birth years, indicating gradual progress in closing urban-rural disparities in female education in Tanzania---though significant differences remain.

\newpage
\subsection{Uncertainty Comparison}
\label{sec:supp-uncertainty-comparison}

Table \ref{tab:admin1-uncertainty-compare} compares the uncertainty in Admin-1 UYS estimates across age groups using three different estimation methods: \textbf{naive weighted estimation}, \textbf{survey GLM}, and \textbf{spatial model}. For each method and age group, the table presents the median width of the 95\% CIs along with the 10th and 90th percentiles of CI width.

\begin{table}[ht]
\centering
\begin{tabular}{lccc}
  \hline
  Age Group & Naive Weighted & Survey GLM & Spatial Model \\ 
  \hline
  15--19  & 1.36 (0.99, 2.65)  & 1.38 (1.04, 2.08)  & 1.17 (0.96, 1.75)  \\ 
  20--24  & 1.88 (1.08, 3.91)  & 1.26 (0.95, 1.88)  & 0.87 (0.77, 1.04)  \\ 
  25--29  & 1.76 (1.17, 2.84)  & 1.26 (0.93, 1.97)  & 0.82 (0.74, 0.99)  \\ 
  30--34  & 2.23 (1.49, 3.18)  & 1.24 (0.93, 1.87)  & 0.85 (0.76, 1.05)  \\ 
  35--39  & 2.25 (1.73, 3.32)  & 1.21 (0.88, 1.81)  & 0.87 (0.78, 1.04)  \\ 
  40--44  & 2.28 (1.76, 3.28)  & 1.18 (0.87, 1.86)  & 0.95 (0.83, 1.16)  \\ 
  45--49  & 2.19 (1.64, 3.12)  & 1.20 (0.86, 1.89)  & 1.11 (0.98, 1.40)  \\ 
  \hline
\end{tabular}
\caption{Comparison of uncertainty in Admin-1 UYS estimates by age group across three estimation methods. Values represent the median width of 95\% CIs, along with the 10th and 90th percentiles.}
\label{tab:admin1-uncertainty-compare}
\end{table}

\begin{figure} [!ht]
    \centering
    \includegraphics[clip, trim=0cm 3.5cm 0cm 3cm, width=0.99\linewidth]{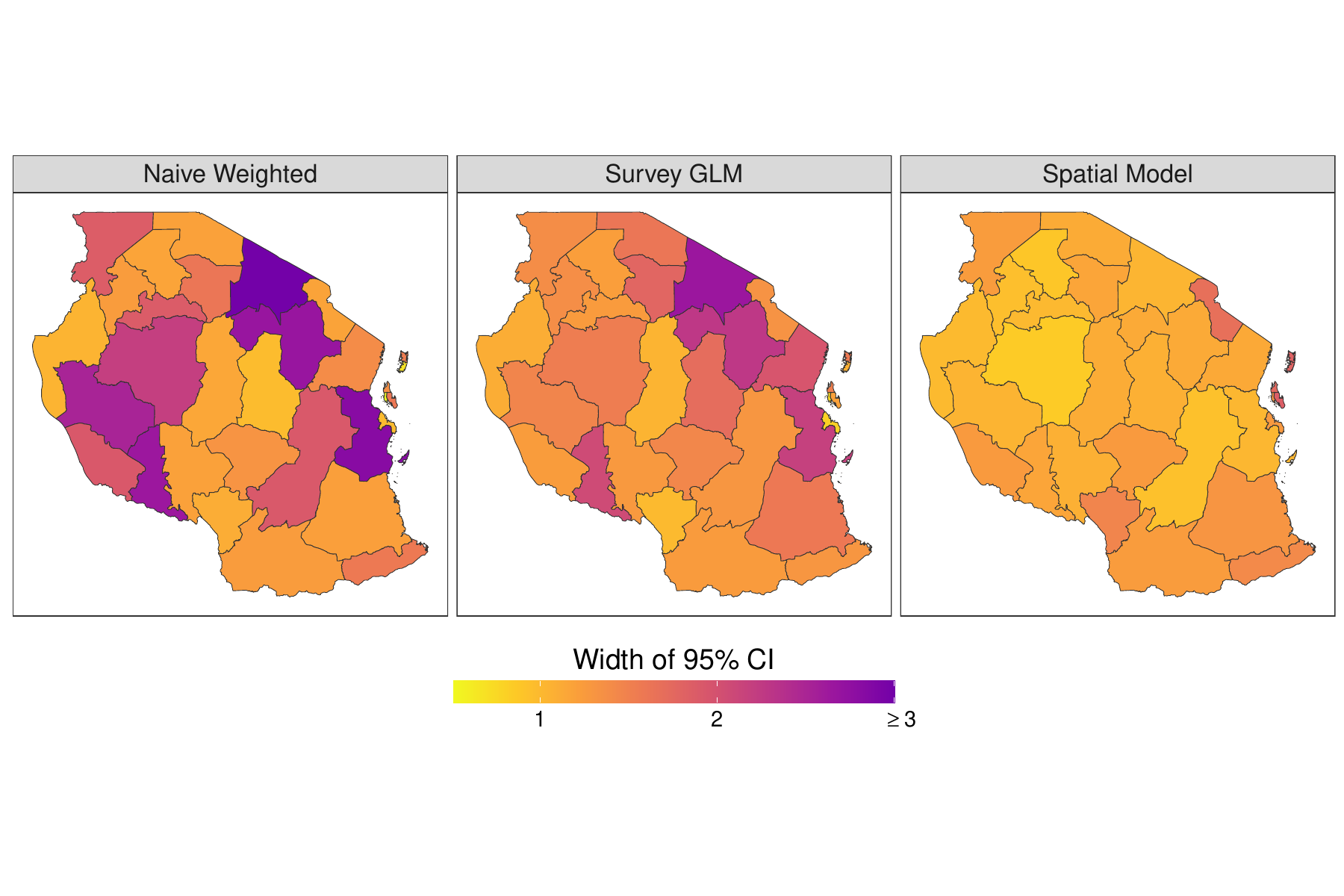}
    \caption{Distribution of widths of 95\% CI for Admin-1 UYS estimates in the 15–19 age group across different estimation methods.}
    \label{fig:adm1-uncertainty-map-15-19}
\end{figure}

Across all age groups, the naive weighted estimates exhibit the widest confidence intervals, indicating greater uncertainty compared to the alternative modeling approaches. The survey GLM method reduces uncertainty substantially, but the spatail model consistently produces the narrowest confidence intervals, indicating the highest level of precision. This pattern is expected, as spatial models impose smoothing structures and borrow information from neighboring domains, leading to more stable estimates. Figure \ref{fig:adm1-uncertainty-map-15-19} illustrates this trend for the 15--19 age group.

\begin{figure} [!ht]
    \centering
    \includegraphics[clip, trim=0cm 1cm 0cm 1cm, width=0.95\linewidth]{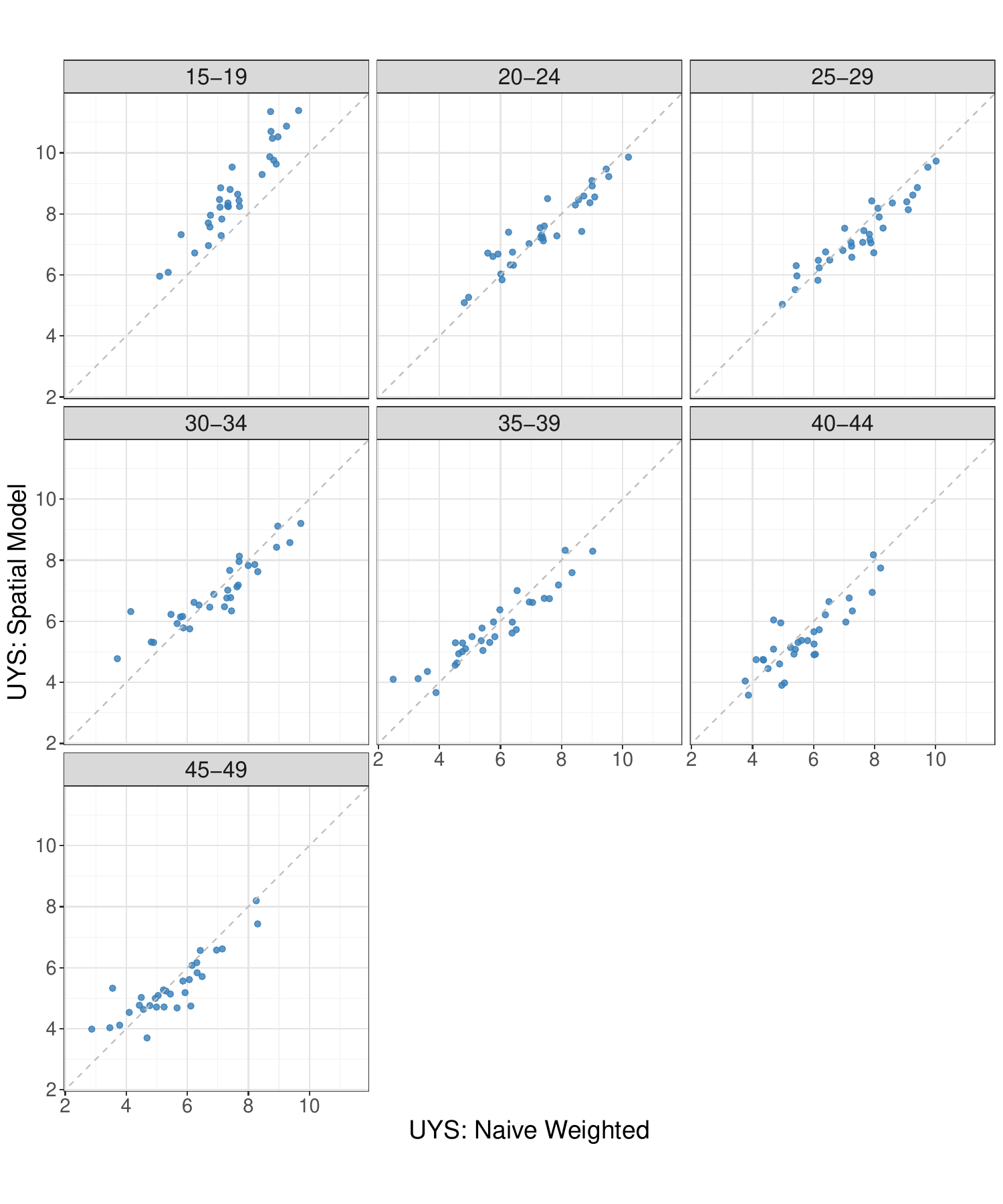}
    \caption{Scatter plot comparing spatial model estimates to naive weighted estimates across age groups and Admin-1 regions.}
    \label{fig:adm1-age-group-direct-vs-INLA}
\end{figure}

To further illustrate the smoothing effect of the spatial model, Figure \ref{fig:adm1-age-group-direct-vs-INLA} presents a scatter plot comparing spatial model estimates to naive weighted estimates. Each point represents a specific combination of age group and Admin-1 region. Compared to the naive weighted estimates, the spatial model estimates tend to be pulled toward the overall mean, reflecting the model's ability to borrow information from neighboring domains and improve precision. Additionally, the top-left panel of the figure shows that naive weighted estimation underestimate UYS for the most recent birth cohorts.

\clearpage
\subsection{Additional Plots on Subnational UYS Estimates}
\label{sec:supp-figure-subnational-UYS}

To further examine subnational educational attainment, we present estimates for females in the 2022 Tanzania DHS, obtained from the spatial model at both Admin-1 and Admin-2 levels. These estimates are based on the framework described in Section \ref{sec:unit-level-model} of the main manuscript.

\begin{figure} [!ht]
    \centering
    \includegraphics[clip, trim=0cm 1.5cm 0cm 1cm, width=0.98\linewidth]{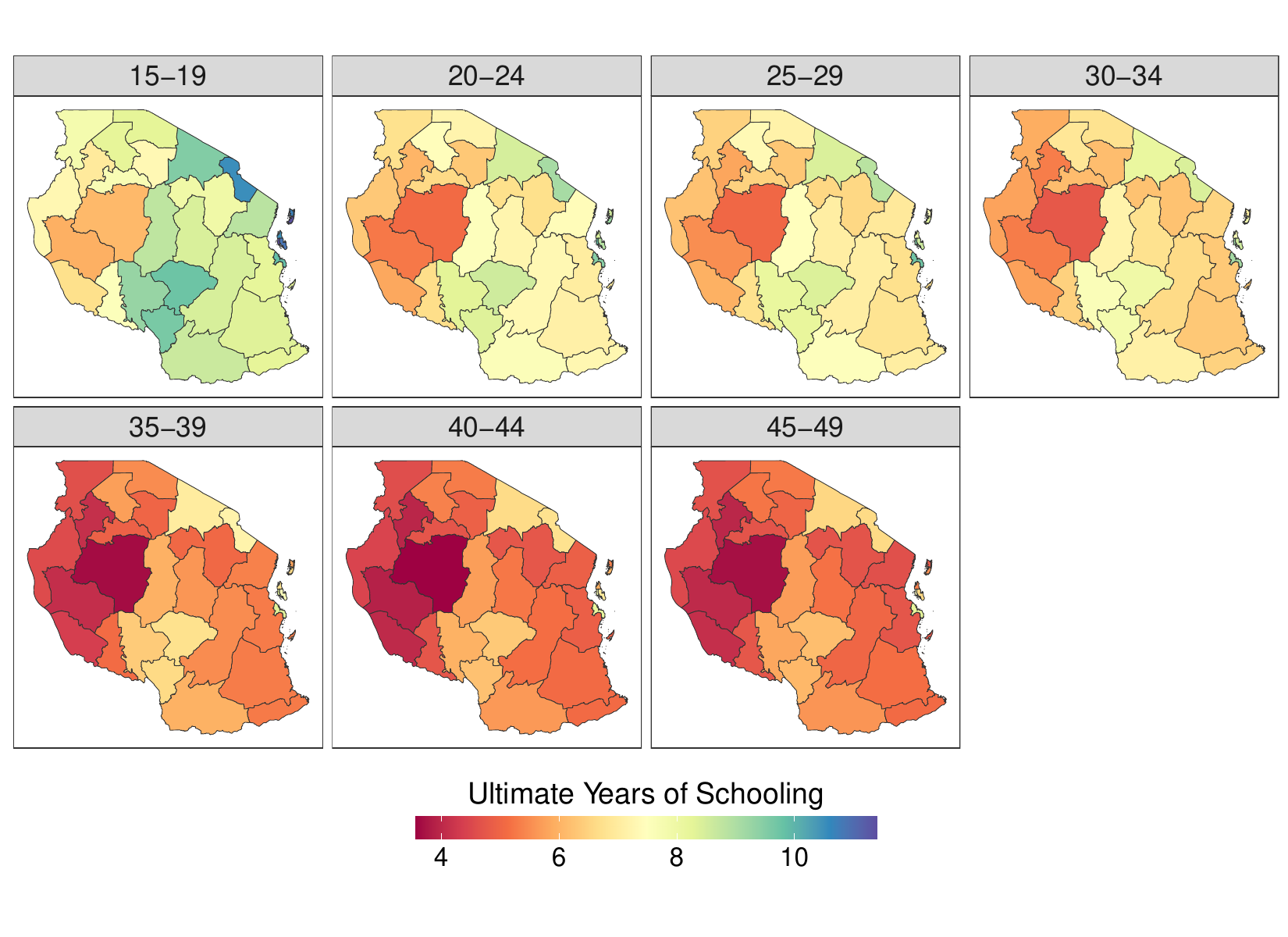}
    \caption{Map for UYS at Admin-1 level across age groups.} 
    \label{fig:TZA-2022-adm1-5yr-map}
\end{figure}

\begin{figure} [!ht]
    \centering
    \includegraphics[clip, trim=0cm 1.5cm 0cm 1cm, width=0.98\linewidth]{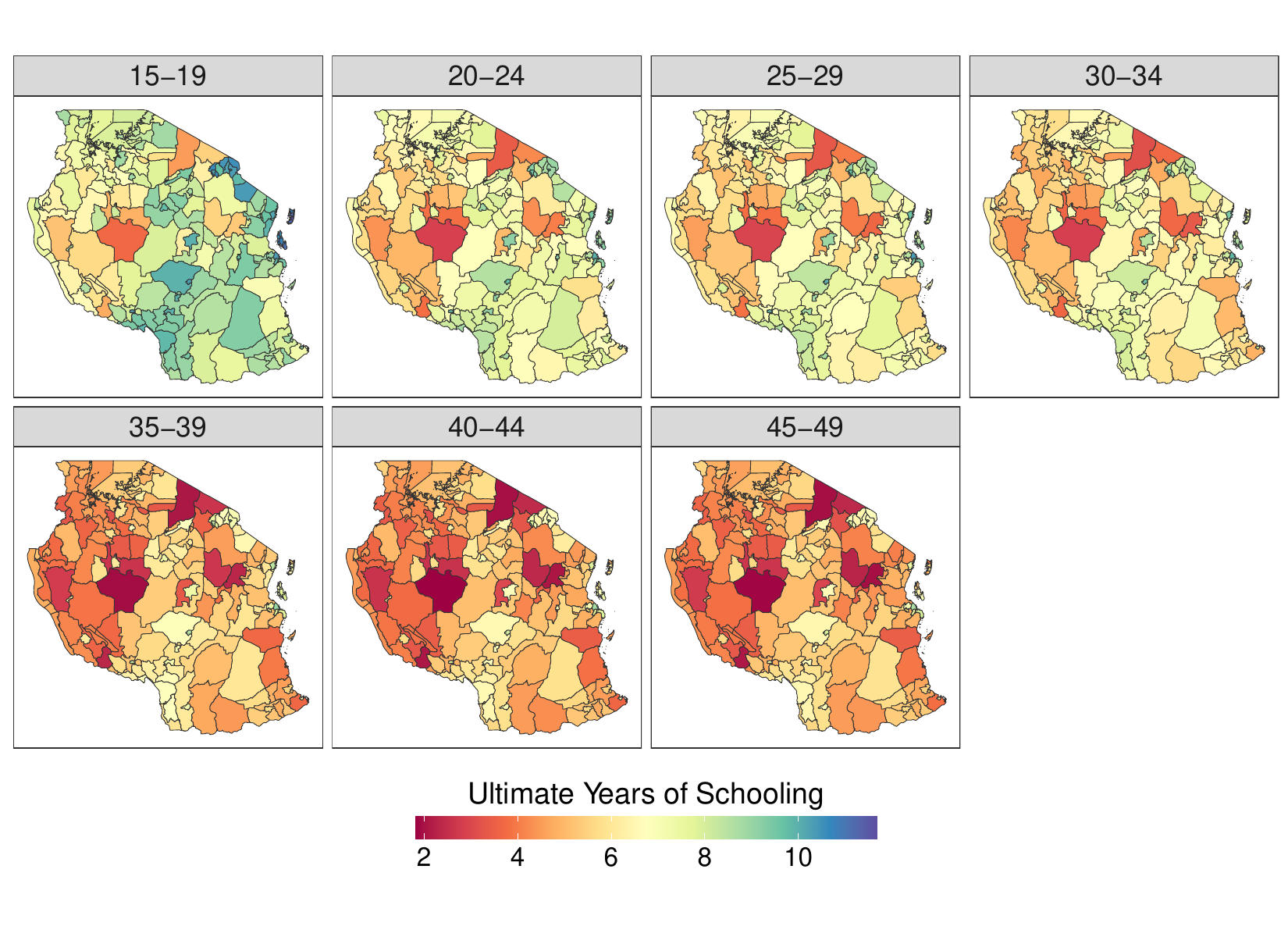}
    \caption{Map for UYS at Admin-2 level across age groups.} 
    \label{fig:TZA-2022-adm2-5yr-map}
\end{figure}

Figures \ref{fig:TZA-2022-adm1-5yr-map} and \ref{fig:TZA-2022-adm2-5yr-map} present the estimated UYS across age groups at Admin-1 and Admin-2 levels, respectively. These maps reveal a general increase in schooling over time, with younger cohorts attaining higher education levels. However, there are strong regional disparities. 

\begin{figure} [!ht]
    \centering
    \includegraphics[clip, trim=0cm 1.5cm 0cm 1cm, width=0.98\linewidth]{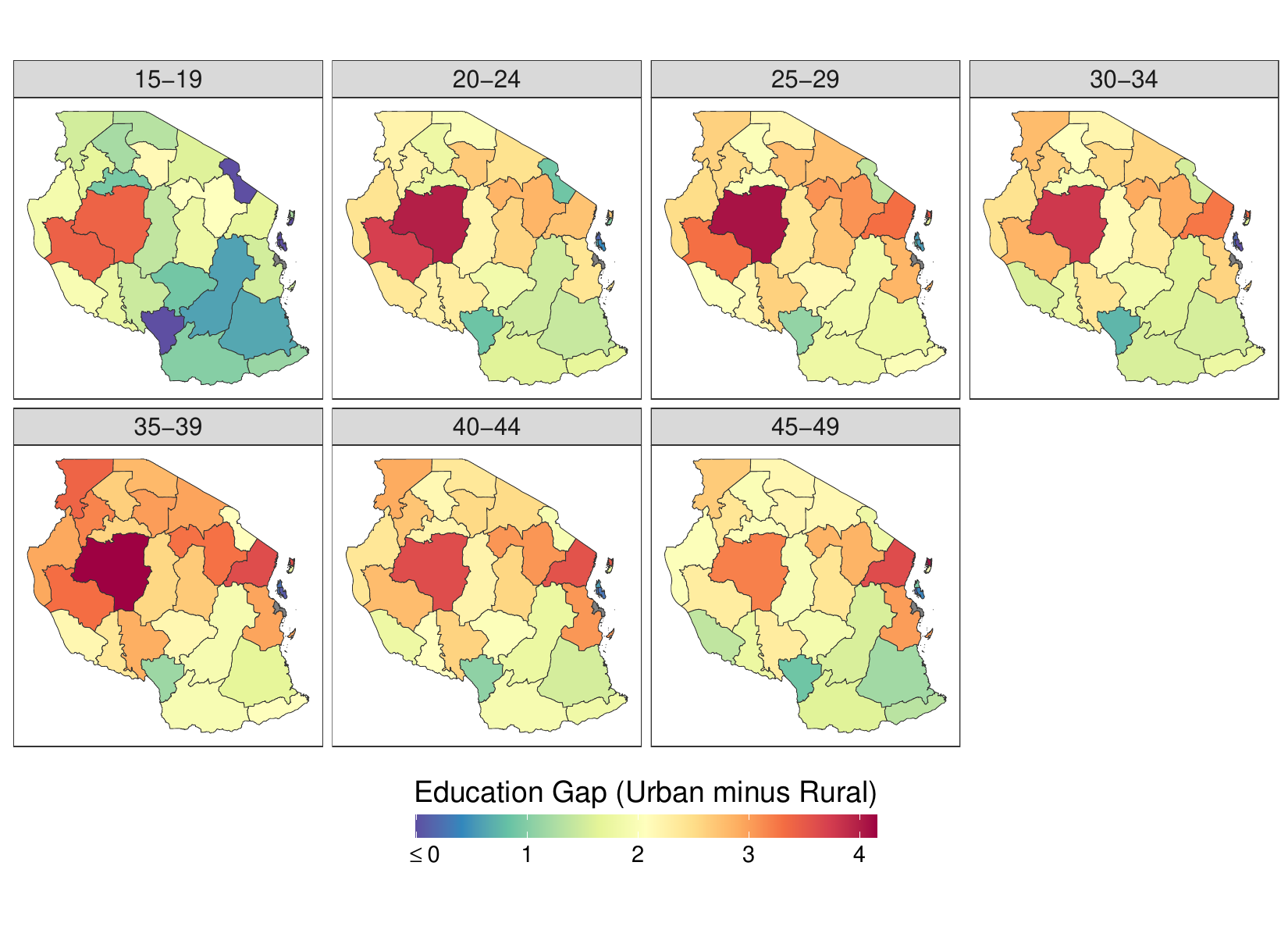}
    \caption{Map for difference in urban and rural UYS (urban-rural) at Admin-2 level across age groups. Dar es Salaam is excluded from the analysis as it is entirely urban. } 
    \label{fig:adm1-agegrp-UR-diff}
\end{figure}

\begin{figure} [!ht]
    \centering
    \includegraphics[clip, trim=0cm 1.5cm 0cm 1cm, width=0.98\linewidth]{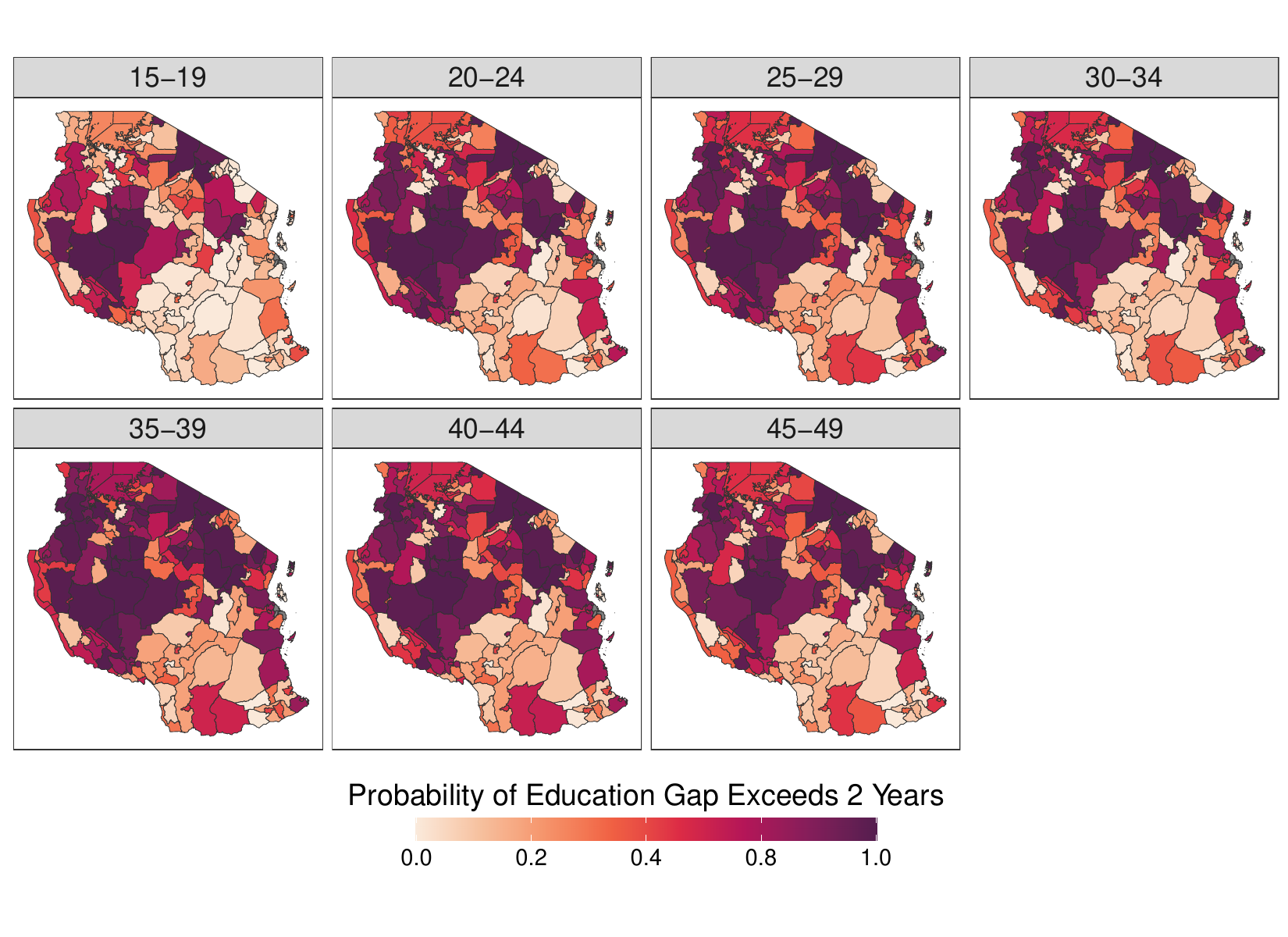}
    \caption{Probability that the urban–rural difference in UYS exceeds two years across Admin-2 regions and age groups. Admin-2 regions within Dar es Salaam is excluded from the analysis as they are entirely urban and do not have rural estimates.}
    \label{fig:adm2-UR-exceedance}
\end{figure}

Figures \ref{fig:adm1-agegrp-UR-diff} and \ref{fig:adm2-UR-exceedance} focus on urban-rural differences. The former shows the mean UYS gap between urban and rural areas at Admin-1 level, highlighting regions where urban female obtain significantly more schooling. The latter presents the probability that the urban-rural gap exceeds two years at Admin-2 level. Notably, Dar es Salaam is excluded as it is entirely urban.

These findings reinforce our conclusion in the main manuscript that subnational and urban/rural disparities in female education are significant, despite overall improvements in educational attainment.

\clearpage
\newpage

\subsection{Additional Metrics for Educational Attainment}
\label{sec:supp-figure-extra-metric}

We present additional statistical metrics naturally derived from our modeling process. Standard survival analysis concepts, such as survival probabilities, have meaningful interpretations in the context of education modeling. Here we present results for female in 2022 Tanzania DHS estimated from the spatial model at Admin-1 level, based on the setup described in Section \ref{sec:unit-level-model} from the main manuscripts. 

\begin{figure} [!ht]
    \centering
    \includegraphics[clip, trim=0cm 0cm 0cm 0cm, width=1.01\linewidth]{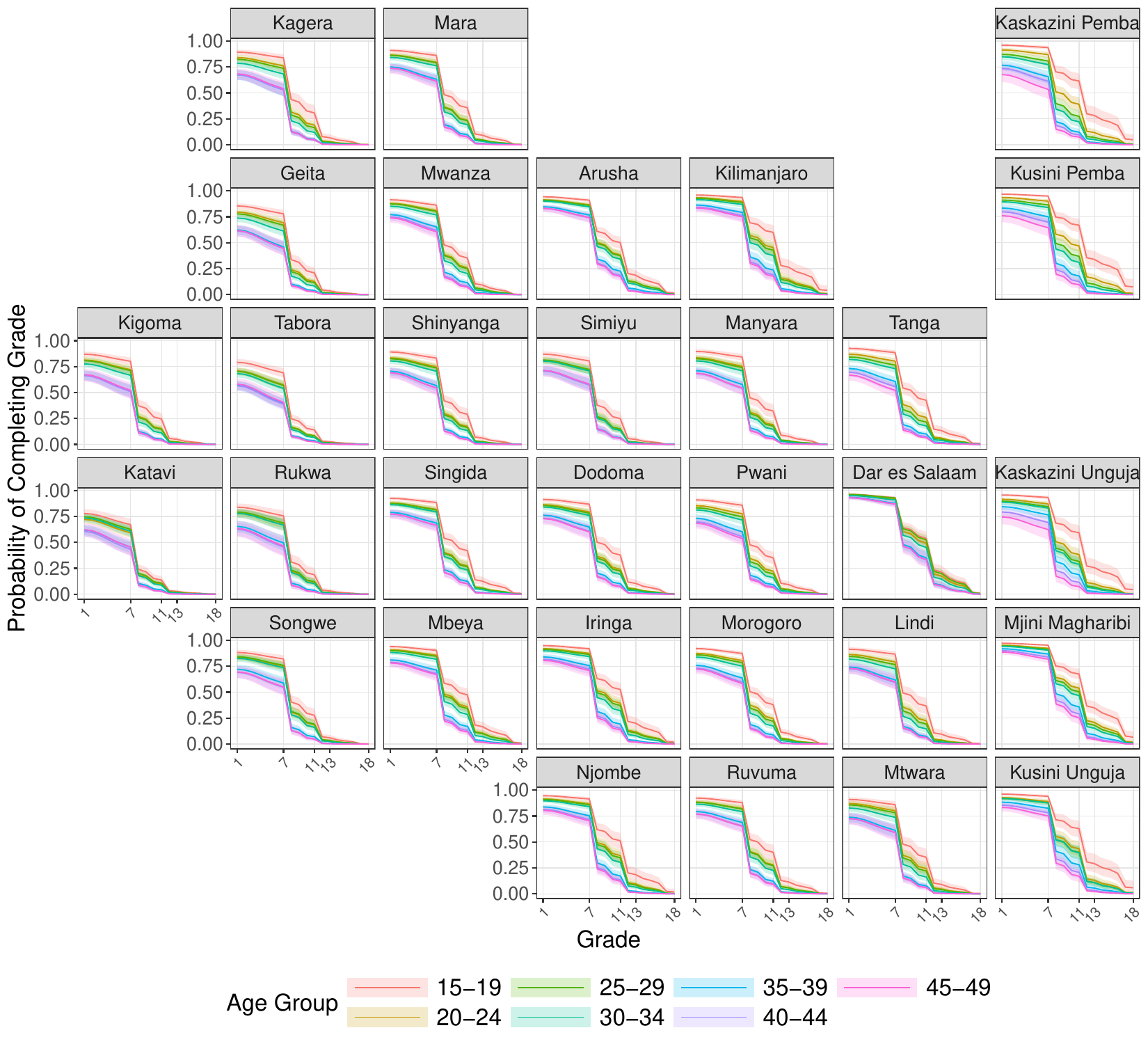}
    \caption{Education progression curves for females across Admin-1 regions in Tanzania, stratified by age group (5-year birth cohorts).}
    \label{fig:TZA-2022-adm1-age-group-surv-geofacet}
\end{figure}

First in Figure \ref{fig:TZA-2022-adm1-age-group-surv-geofacet}, we examine the marginal probability of reaching each grade, i.e., $S(t)=P(T\ge t)$ for $t \in [1,18]$.  This metric depicts the average educational trajectory for a group, with each curve representing females from an Admin-1 region within a given birth cohort. Key educational milestones---school entry (Grade 1), primary school completion (Grade 7), lower secondary (Grade 11), and upper secondary (Grade 13)---are marked along the x-axis. Substantial dropout rates are observed at transitions between education levels, with the most pronounced declines occurring after primary and lower secondary school.

\begin{figure} [!ht]
    \centering
    \includegraphics[clip, trim=0cm 0cm 0cm 0cm, width=1.01\linewidth]{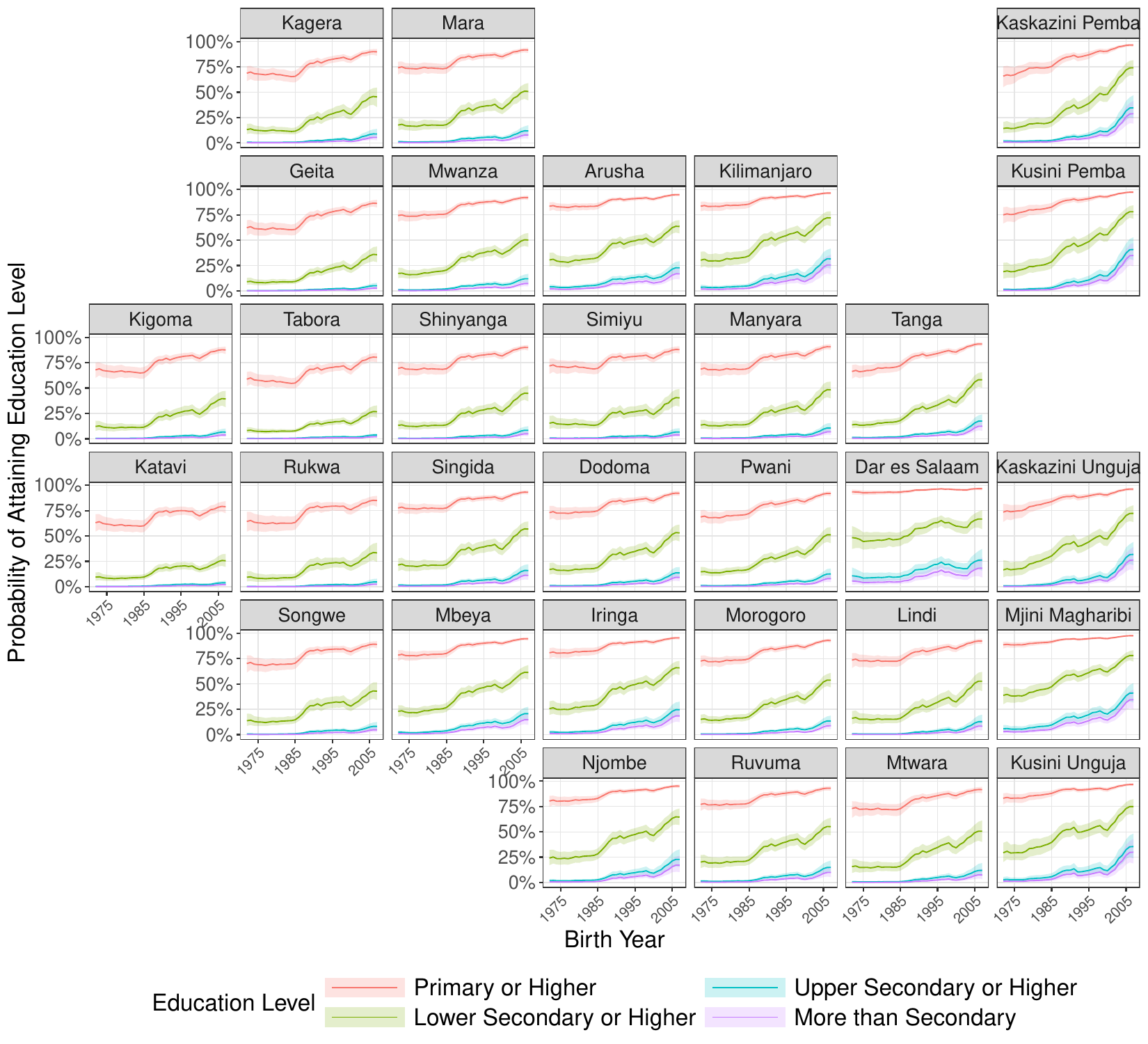}
    \caption{Probability of attaining at least primary, lower secondary, upper secondary, or post-secondary education for females across Admin-1 regions in Tanzania, stratified by age group (5-year birth cohort).}    
    \label{fig:TZA-2022-adm1-yearly-highest-attained-geofacet}
\end{figure}

Beyond grade-wise progression, Figure \ref{fig:TZA-2022-adm1-yearly-highest-attained-geofacet} presents the highest level of education attained across regions and birth cohorts. The increasing trend highlight substantial improvements in education attainment for all levels over time, but regional disparities persist.

\begin{figure} [!ht]
    \centering
    \includegraphics[clip, trim=0cm 0cm 0cm 0cm, width=1.01\linewidth]{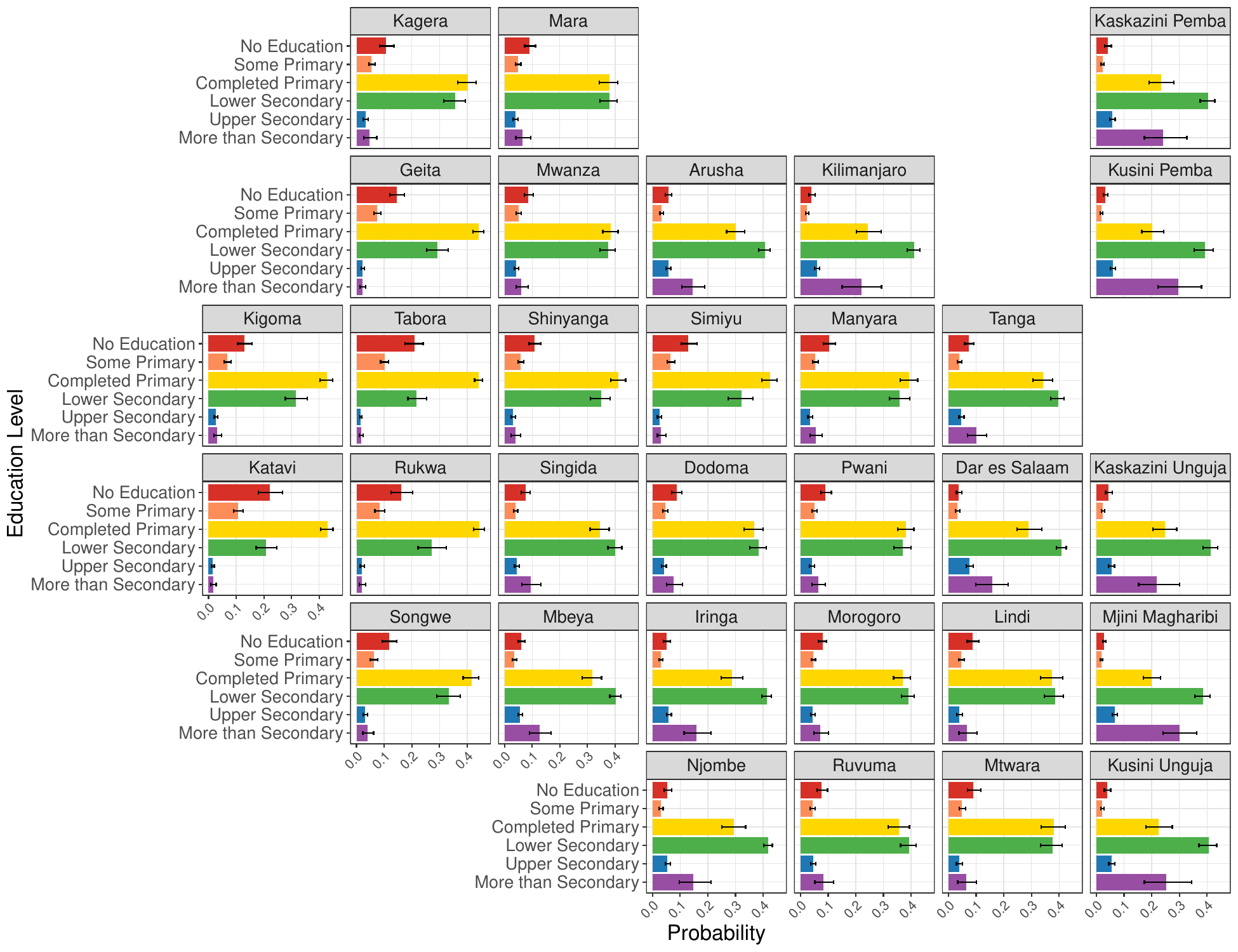}
    \caption{Distribution of (ultimate) education levels across Admin-1 regions for females aged 15--19 in 2022, with 95\% credible intervals.}
    \label{fig:TZA-2022-adm1-15-19-level-dist-geofacet}
\end{figure}

Finally, Figure \ref{fig:TZA-2022-adm1-15-19-level-dist-geofacet} exhibits the estimated distribution of (ultimate) education levels among females aged 15--19 in the 2022 Tanzania DHS. In urban areas like Dar es Salaam, females are more likely to attain higher education levels, whereas rural regions have a larger proportion with incomplete primary education or no education. These patterns highlight persistent and substantial  disparities in educational attainment.

\newpage

\newpage

\end{document}